\documentclass[%
 reprint,
nofootinbib,
 amsmath,amssymb,
 aps,
]{revtex4-2}

\usepackage{graphicx}
\usepackage{dcolumn}
\usepackage{bm}

\usepackage{feynmp}
\usepackage{tikz-feynman}
\tikzfeynmanset{compat=1.1.0}
\usetikzlibrary{shapes,arrows,positioning,automata,backgrounds,calc,er,patterns}

\usepackage{amsmath}
\usepackage{amssymb}
\usepackage[colorlinks=true,citecolor=magenta,urlcolor=cyan]{hyperref}
\usepackage{multirow}

\usepackage{lineno}

\def\Hcoll{H_{1q}^{\perp\, h}}
\def\Hinterference{H_{1q}^{\sphericalangle\, hh}}

\def\antiq{\bar{q}_1}
\def\q{q_1}
\def\qp{q_2}
\def\h{h}
\def\sh{s_{h}}
\def\qbar{\bar{q}_1}
\def\qpbar{\bar{q}_2}
\def\hbar{H}

\def\pElec{\textbf{p}_-}

\def\kQuark{\textbf{k}_{1}}
\def\kAntiQuark{\bar{\textbf{k}}_{1}}

\def\betaq{\beta_q}

\def\eh{\varepsilon_{\h}}
\def\ehbar{\varepsilon_{\hbar}}
\def\ql{q_{l}}
\def\qlbar{\bar{q}_{n}}

\def\helElec{\lambda_-}
\def\helPos{\lambda_+}
\def\helQuark{\lambda_{1}}
\def\helQuarkp{\lambda_{2}}
\def\helAntiQuark{\bar{\lambda}_{1}}
\def\helAntiQuarkp{\bar{\lambda}_{2}}
\def\M{\hat{\mathcal{M}}}

\def\C{\mathcal{C}}
\def\sigmaq{\sigma^{\q}}
\def\sigmaqbar{\sigma^{\antiq}}
\def\Iq{1^{q_1}}
\def\Iqp{1^{\qp}}
\def\sigmaXq{\sigma_x^{\q}}
\def\sigmaYq{\sigma_y^{\q}}
\def\sigmaZq{\sigma_z^{\q}}
\def\Iqbar{1^{\qbar}}

\def\sigmaXqbar{\sigma_x^{\antiq}}
\def\sigmaYqbar{\sigma_y^{\antiq}}
\def\sigmaZqbar{\sigma_z^{\antiq}}
\def\Iden{1}

\def\GeV{\rm{GeV}}

\def\zu{\hat{\textbf{z}}}
\def\xq{\hat{\textbf{x}}_{\q}}
\def\yq{\hat{\textbf{y}}_{\q}}
\def\zq{\hat{\textbf{z}}_{\q}}
\def\xqbar{\hat{\textbf{x}}_{\antiq}}
\def\yqbar{\hat{\textbf{y}}_{\antiq}}
\def\zqbar{\hat{\textbf{z}}_{\antiq}}

\def\QHF{\rm{QHF}}
\def\AHF{\rm{AHF}}

\def\p{p}
\def\k{k_{1}}
\def\kp{k_{2}}
\def\pt{\textbf{p}_{\rm T}}
\def\Pt{\textbf{P}_{\rm T}}
\def\kT{\textbf{k}_{\rm T}}
\def\kTkT{\rm{k}^2_{\rm T}}

\def\kt{\textbf{k}_{1\rm T}}
\def\kpt{\textbf{k}_{2\rm T}}

\def\klt{\textbf{k}_{l\rm T}}
\def\kltbar{\bar{\textbf{k}}_{n\rm T}}
\def\kptkpt{\textbf{k}^2_{2\rm T}}
\def\ptpt{\textbf{p}^2_{\rm T}}
\def\PtPt{\textbf{P}^2_{\rm T}}
\def\ktabs{\rm{k}_{1\rm T}}
\def\kptabs{\rm{k}_{2\rm T}}
\def\kbar{\bar{k}_1}
\def\kpbar{\bar{k}_2}
\def\ktbar{\bar{\textbf{k}}_{1\rm T}}
\def\kptbar{\bar{\textbf{k}}_{2\rm T}}

\def\kptbarkptbar{\bar{\textbf{k}}^2_{2\rm T}}

\def\kptabsbar{\bar{k}_{2\rm T}}
\def\mh{M}
\def\bL{b_{\rm L}}
\def\bT{b_{\rm T}}
\def\fT{f_{\rm T}}

\def\kptilde{\tilde{\textbf{k}}_{2\rm T}}

\def\i{\rm i}
\def\sigmaqT{\boldsymbol{\sigma}_{\rm T}}

\def\GL{G_{\rm L}}
\def\GT{G_{\rm T}}
\def\V{\textbf{V}}
\def\VL{V_{\rm L}}
\def\VT{\textbf{V}_{\rm T}}
\def\fVM{f_{\rm VM}}
\def\fL{f_{\rm L}}
\def\thetaLT{\theta_{\rm LT}}
\def\Gammaq{\Gamma}
\def\Gammaqbar{\Gamma}
\def\GammaqDag{\Gamma^{\dagger}}
\def\GammaqbarDag{\Gamma^{\dagger}}
\def\PS{\rm{PS}}
\def\VM{\rm{VM}}
\def\Cqq{\C^{\q\antiq}}
\def\phikptQ{\phi_{\textbf{k}_2}^{\q}}
\def\phikptQbar{\phi_{\bar{\textbf{k}}_2}^{\qbar}}
\def\Trqq{\rm{Tr}_{\q\qbar}}
\def\Tr{\rm{Tr}}
\def\Trq{\rm{Tr}_{q}}

\def\Sq{\textbf{S}_{\q}}
\def\SqT{\textbf{S}_{\q\rm T}}
\def\SqL{\textbf{S}_{\q\rm L}}
\def\phiSq{\phi_{\Sq}}

\def\Ptot{P_{\rm tot}}
\def\PtotT{\textbf{P}_{\rm tot,\rm T}}
\def\T{\textbf{T}}

\def\be{\begin{equation}}
\def\ee{\end{equation}}

\def\la{\langle}
\def\ra{\rangle}

\def\Sv{{\bf S}}
\def\pv{{\bf{p}}}

\begin{document}

\preprint{APS/123-QED}

\title{String fragmentation of a quark pair with entangled spin states:\\ application to $e^+e^-$ annihilation}

\author{A. Kerbizi$^{a}$}\email{albi.kerbizi@ts.infn.it}
\author{X. Artru$^{b}$} \email{x.artru@ipnl.in2p3.fr}
\affiliation{
$^{a}$Dipartimento di Fisica, Universit\`a  degli Studi di Trieste and INFN Sezione di Trieste,\\
 Via Valerio 2, 34127 Trieste, Italy\\ 
$^{b}$Universit\'e de Lyon, Institut de Physique des deux Infinis (IP2I Lyon), Universit\'e Lyon 1 and CNRS,\\
 4 rue Enrico Fermi, F-69622 Villeurbanne, France \\
}%

\date{\today}

\begin{abstract}
We present a recursive quantum mechanical model for the fragmentation of a string stretched between a quark and an antiquark with entangled spin states. The quarks are assumed to be produced in the $e^+e^-$ annihilation process via the exchange of a virtual photon and the correlations between their spin states are described by a joint spin density matrix. The string fragmentation process is formulated at the amplitude level by using the splitting matrices of the recent string+${}^3P_0$ model of polarized quark fragmentation with pseudoscalar and vector meson emissions, and accounts for the systematic propagation of the spin correlations in the fragmentation chain. The model is formulated as a recursive recipe suitable for a Monte Carlo implementation. It reproduces the expected angular correlation, due to the Collins effect, between back-to-back pseudoscalar and/or vector mesons. For the latter, this correlation also involves the momenta of the decay products. We use the model for studying the sign of the Collins asymmetry for back-to-back vector and pseudoscalar mesons. \end{abstract}

\maketitle


\section{Introduction}
The $e^+e^-$ annihilation to hadrons is an important process to study the hadronization of quarks and gluons in the observed hadrons, a still poorly understood phenomenon of the strong interactions. According to the factorization theorem in quantum chromodynamics (QCD) \cite{Collins:1981uk}, the cross section can be factorized in the cross section for the annihilation reaction $e^+e^-\rightarrow \q\,\qbar$ and in the fragmentation functions (FFs) that describe the conversion of the quark $\q$ and antiquark $\qbar$ in the observed hadrons. Recently the $e^+e^-$ annihilation has been used as a tool to access the class of spin-dependent FFs. Examples are the Collins function $\Hcoll$ \cite{Collins:1992kk}, which describes the fragmentation of a transversely polarized quark $\q$ in an unpolarized hadron $h$, and the interference fragmentation function (IFF) $\Hinterference$ \cite{Collins:1993kq,Bianconi:1999cd}, which describes the fragmentation of a transversely polarized quark $\q$ in a pair of unpolarized hadrons $hh$.

In $e^+e^-$ annihilation cross section, the spin-dependent FFs give birth to azimuthal modulations in the distribution of the observed hadrons. The amplitudes of these modulations are referred to as asymmetries and provide access to the spin-dependent FFs (for a review see e.g. Ref. \cite{Boer:2008fr}). The semi-inclusive annihilation process $e^+e^-\rightarrow h_1h_2X$, where one of the hadrons $h_1$ or $h_2$ is assumed to be produced in the quark jet and the other in the antiquark jet, allows to measure the Collins asymmetry originated from the coupling of two Collins FFs \cite{Boer:2008fr}. If two hadrons are observed in each quark jet by the process $e^+e^-\rightarrow (h_1h_2)(\bar{h}_1\bar{h}_2)X$, the Artru-Collins asymmetry \cite{Artru:1995zu} appears, which allows to access the product of two IFFs. The Collins asymmetries have been measured in $e^+e^-$ by the BELLE \cite{Belle:2008fdv,Belle:2019nve}, BABAR \cite{BaBar:2013jdt,BaBar:2015mcn} and BESIII \cite{BESIII:2015fyw} experiments, whereas the Artru-Collins asymmetries have been measured by the BELLE experiment \cite{Belle:2011cur}.

The $e^+e^-$ asymmetry data have played a fundamental role in the investigation of the partonic structure of the nucleons. The data have been analyzed in combination with the data from semi-inclusive deep inelastic scattering (SIDIS) of leptons off transversely polarized nucleons on the Collins asymmetries \cite{Collins:1992kk} and on the dihadron production asymmetry \cite{Collins:1993kq,Bianconi:1999cd} to extract the spin-dependent FFs and the transversity parton distribution function (see e.g. Refs. \cite{Anselmino:2007fs,Martin:2014wua,Radici:2015mwa,Kang_transv_evolution,Cocuzza:2023oam}). Transversity describes the transverse polarization of quarks in a transversely polarized nucleon, and is the third parton distribution function needed to characterize the collinear partonic structure of the nucleon at leading order.

This work is dedicated to the modeling of spin-effects in $e^+e^-$ annihilation to hadrons with the final goal of implementing the model in a Monte Carlo event generator (MCEG). Recently, the spin effects have been implemented for the simulation of the polarized SIDIS process in the PYTHIA 8 MCEG \cite{Sjostrand:2007gs,Bierlich:2022pfr} by the StringSpinner package \cite{Kerbizi:2021StringSpinner,Kerbizi:2023cde}. The simulation of the spin effects is based on the string+${}^3P_0$ model of polarized hadronization \cite{Kerbizi:2018qpp,Kerbizi:2019ubp,Kerbizi:2021M20}, which is implemented in StringSpinner. It is a quantum mechanical extension of the Lund Model \cite{Andersson:1983ia} of string fragmentation that includes the quark spin degree of freedom at the amplitude level. An analogue MCEG for $e^+e^-$ annihilation with spin effects presently does not exist, the main difficulty coming from the non-classical (entangled) correlation of the $\q$ and $\qbar$ spins.

Up to now, the string+${}^3P_0$ model has been applied to the description of the recursive fragmentation of a string, starting from one end drawn by a polarized quark or antiquark without taking care of the polarization of the object (quark or diquark) drawing the opposite end. Thus it can not be applied as it stands to $e^+e^-$ annihilation where both $\q$ and $\qbar$ are polarized, which is more in an entangled fashion \cite{Chen:1994ar}. Here we extend the string+${}^3P_0$ model to the description of the fragmentation of a string stretched between a quark $\q$ and an antiquark $\qbar$ with entangled spin states. The correlation between their spin states is described by the means of a joint spin density matrix of the pair. The joint spin density matrix is calculated assuming the $e^+e^-$ annihilation to be mediated by a virtual photon. 
The model is however general and it does not depend on the mechanism invoked for the production of the $\q\qbar$ pair. The rules of the string+${}^3P_0$ model with emission of pseudoscalar (PS) mesons and vector mesons (VMs) \cite{Kerbizi:2021M20} are used to describe the hadron emissions from the quark and antiquark ends of the string. To take into account the quantum mechanical correlations between the two endpoints of the string after hadron emissions we employ a recipe inspired from the Collins-Knowles (CK) recipe \cite{Collins:1987cp,Knowles:1988vs,Richardson:2001df}. The CK recipe is applied to involve in the correlation not only the momentum of a VM but also the individual momenta of its decay products. Finally we formulate a recursive recipe for the simulation of the string fragmentation of a $\q\qbar$ pair with correlated spin states that is suitable for a Monte Carlo implementation. The recipe is applied to the production of the two leading hadrons in $e^+e^-\rightarrow h_1\,h_2X$ showing that it reproduces the expected modulations in the azimuthal distribution of the hadrons \cite{Boer:2008fr,DAlesio:2021dcx}.

Throughout the article we neglect gluon emissions from the quark and the antiquark. An extension of the model to the introduction of pQCD effects appears not to be straightforward, and it will be considered in a future work.

The article is organized as follows. In Sec. \ref{sec:ingredients} we give the main ingredients needed for the modeling of the spin effects in $e^+e^-$ annihilation. In Sec. \ref{sec:polarized string fragmentation} the different steps needed to describe the fragmentation of a string stretched between a quark pair with entangled spin state are discussed and the final recursive recipe suitable for Monte Carlo implementation is given. The recipe is applied to the production of two back-to-back hadrons in Sec. \ref{sec:application}. Finally the conclusions are drawn in Sec. \ref{sec:conclusions}.
\vspace{3em}

\section{Physics ingredients for modeling spin effects in $e^+e^-$ annihilation}\label{sec:ingredients}
Following the factorization theorem \cite{Collins:1981uk}, we factorize the annihilation process $e^+e^-\rightarrow \q\antiq\rightarrow h_1,h_2,\dots,h_N$, where $h_1,h_2,\dots,h_N$ are the final state hadrons, into a hard sub-process where the quark pair $\q\qbar$ is created in the annihilation $e^+e^-\rightarrow \q\qbar$, and a soft process $\q\qbar\rightarrow h_1,h_2,\dots,h_N$, where the quark pair hadronizes into the final state hadrons. The framework for the description of the hard subprocess is set up in Sec. \ref{sec:kinematics} from the kinematical point of view, and in Appendix \ref{sec:helicity amplitudes} from the dynamical point of view.

We describe the hadronization of the $\q\qbar$ pair by an extension of the string +${}^3P_0$ model to the fragmentation of a string stretched between $\q$ and $\qbar$, which have correlated spin states. To this end in Sec. \ref{sec:unitarity diagram} we introduce the folded unitarity diagram (Fig. \ref{fig:amplitude squared e+e-}) for the process $e^+e^-\rightarrow \q\qbar\rightarrow h\,H\,X$, where $h$ and $H$ are hadrons associated to the $\q$ and $\qbar$ jets, respectively. The diagram is related to the cross section of the process, which depends on the joint spin density matrix of the $\q\qbar$ pair and on the splitting amplitudes for the recursive emissions of hadrons from the quarks. The joint spin density matrix of the $\q\qbar$ pair is described in Sec. \ref{sec:joint spin density matrix}, while the splitting amplitudes, which are taken from the string+${}^3P_0$ model of Ref. \cite{Kerbizi:2021M20}, are recalled in Sec. \ref{sec:hadronization model}.


\subsection{The unitarity diagram for $e^+e^-$ annihilation}\label{sec:unitarity diagram}
\begin{figure}[tbh]
\centering
\begin{minipage}[b]{0.5\textwidth}
\includegraphics[width=0.9\textwidth]{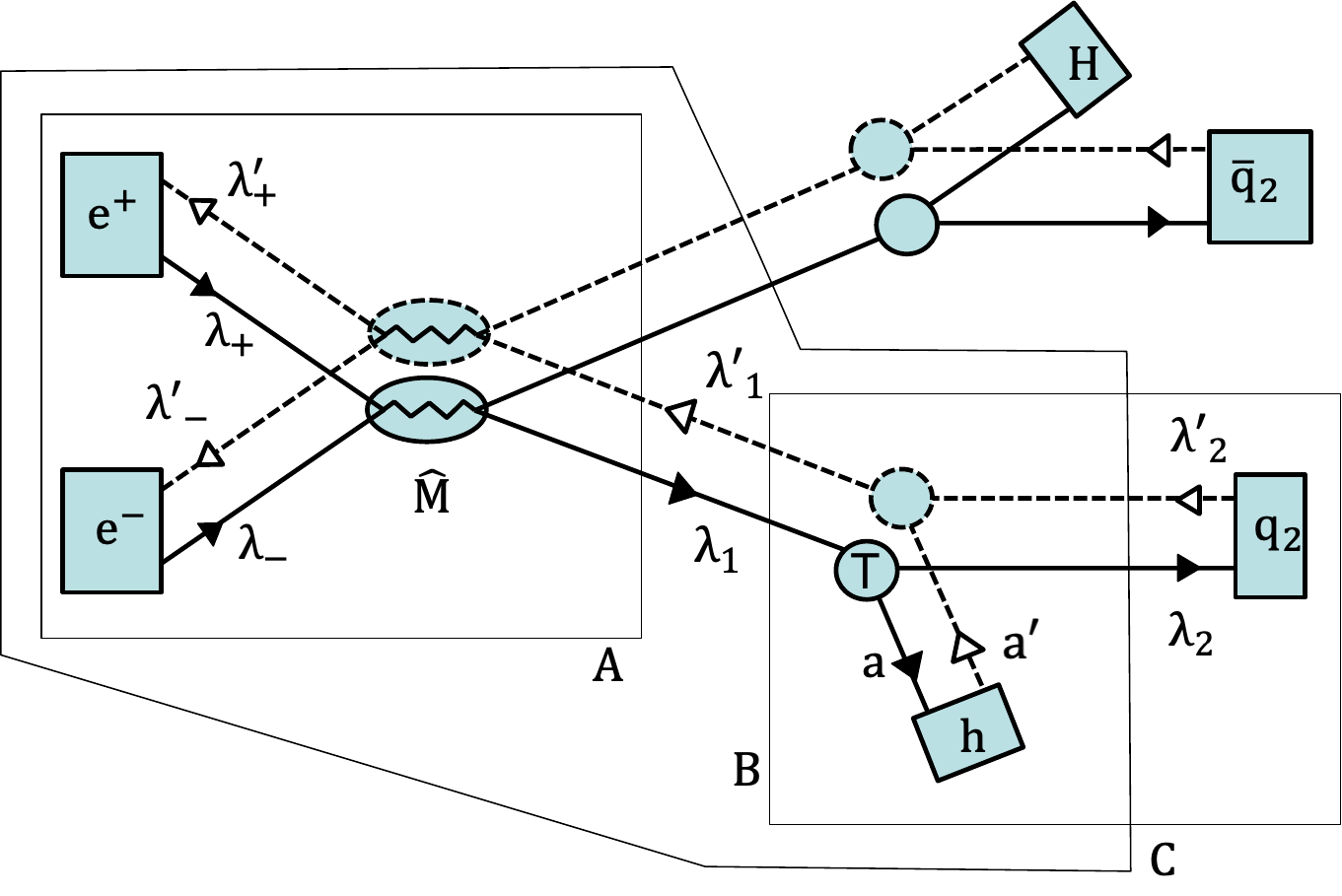}
\end{minipage}
\caption{Folded unitarity diagram for the reaction $e^+\!\! \uparrow e^-\!\!\uparrow \ \to \q\qbar$ followed by the splittings $\q \to h+X$ and $\qbar \to H+X'$.}
\label{fig:amplitude squared e+e-}
\end{figure}

\textbf{Conventions for the figures}. We represent the probability of a reaction by a pair of Feynman-like diagrams, slightly shifted from each other: a direct diagram with black arrow-points for the amplitude and a reversed diagram with white arrow-points for the complex conjugate amplitude (see, e.g., Fig. \ref{fig:amplitude squared e+e-}). This representation is obtained by folding the standard, left-right symmetrical, unitarity diagram. Gray blobs represent subprocess amplitudes.	
Greek letters label the spin states of fermions, latin letters those of bosons. These letters are primed for the conjugate amplitude. 

Gray rectangles represent sources of initial particles or detectors of final particles. A source is characterized by a density or {\it emittance}\footnote{If not normalized to unit trace.} 
matrix $\langle\lambda|\rho|\lambda'\rangle$, a detector by an {\it acceptance matrix} \cite{Artru:2008cp} $\langle\lambda'|\eta|\lambda\rangle$. For an unpolarized source (resp. detector), the $\delta_{\lambda,\lambda'}$ is represented by a U-turn 
$\subset_{\!\blacktriangleright\!-}^{\!\triangleleft-}$ 
(resp. $_{-\!\blacktriangleright}^{-\!\triangleleft}\!\!\!\supset$)
of the particle line inside the gray rectangle. Also, the unit acceptance matrix is indicated by $\eta^{\rm U}$ in the figures.

\textbf{Elements of the unitarity diagram.} To introduce the main ingredients required for the description of $e^+e^-$ annihilation, we start by the folded unitarity diagram for the process $e^+e^-\rightarrow \q\qbar\rightarrow h\,H\,X$ shown in Fig. \ref{fig:amplitude squared e+e-}. The hadron $h$ is emitted in the quark splitting $\q\rightarrow h + \qp$. The hadron $H$ is emitted in the antiquark splitting $\qbar\rightarrow H+\qpbar$. The fragmentation chains initiated by the left-over quarks $\qp$ and $\qpbar$ are not shown explicitly, but summarized by acceptance matrices represented by gray boxes.

To the folded unitarity diagram is associated the two-particle inclusive cross section
\begin{equation}\label{eq:A^2 e+e-}
\begin{aligned}
    &d\sigma (e^+\!\! \uparrow e^-\!\! \uparrow\rightarrow h\,H\,X) \propto \langle \helQuarkp|T_{\qp,h,\q}|\helQuark\rangle \, \langle \helAntiQuarkp |T_{\qpbar,H,\qbar}|\helAntiQuark\rangle\\
    &\times \langle\helQuark,\helAntiQuark|\M|\helElec,\helPos\rangle\,\rho_{\helElec\helElec'}^{e^-}\,\rho_{\helPos\helPos'}^{e^+}\, \langle \helElec',\helPos'|\M^{\dagger}|\helQuark',\helAntiQuark'\rangle \\
    & \times\langle \helQuark'|T^{\dagger}_{\qp,h,\q}|\helQuarkp'\rangle\,\langle \helAntiQuark'|T^{\dagger}_{\qpbar,H,\qbar}|\helAntiQuarkp'\rangle \\
    &\times \langle \helQuarkp'|\eta(\qp)|\helQuarkp\rangle\,\langle \helAntiQuarkp'|\eta(\qpbar)|\helAntiQuarkp\rangle,
\end{aligned}
\end{equation}
where the repeated indices are summed over. The second line in Eq. (\ref{eq:A^2 e+e-}) expresses the production of the $\q\qbar$ pair from the annihilation of the $e^+$ and $e^-$. It describes the hard subprocess $e^+e^-\rightarrow \q\qbar$. The first and third lines describe the splittings $\q\rightarrow h +\qp$ and $\qbar\rightarrow H +\qpbar$ in the amplitude and the complex conjugated amplitude, respectively. The last line expresses possible further information coming from the successive splittings of $\qp$ and $\qpbar$. The meaning of the different components of the amplitude is as follows.

The quantity $\M$ indicates the quantum amplitude associated to the hard process $e^+(\helPos)\,e^-(\helElec)\rightarrow \q(\helQuark)\,\qbar(\helAntiQuark)$. We have labelled with $\helPos$, $\helElec$, $\helQuark$ and $\helAntiQuark$ the helicities of $e^+$, $e^-$, $\q$ and $\qbar$, respectively. The quantum states $|\helElec,\helPos\rangle$ and $|\helQuark,\helAntiQuark\rangle$ indicate respectively the two-particle helicity states $|\helElec\rangle\otimes |\helPos\rangle$ and $|\helQuark\rangle \otimes |\helAntiQuark\rangle$. $\M_{\helElec\helPos;\helQuark,\helAntiQuark}\equiv\langle \helQuark, \helAntiQuark|\M|\helElec,\helPos\rangle$ is the helicity amplitude, which can be calculated perturbatively using standard methods. The expressions that we obtain are given for completeness in Appendix \ref{sec:helicity amplitudes}.

The quantities $\rho^{e^-}_{\helElec\helElec'}$ and $\rho^{e^+}_{\helPos\helPos'}$, represented by the gray boxes on the left of Fig. \ref{fig:amplitude squared e+e-}, are the spin density matrices of the $e^-$ and $e^+$, respectively. They contain the information on the spin states of the beam particles. In the following we assume the electron and positron to be unpolarized, i.e. $\rho^{e^-}_{\helElec\helElec'}=\delta_{\helElec\helElec'}/2$ and $\rho^{e^+}_{\helPos\helPos'}=\delta_{\helPos\helPos'}/2$.

The operator $T_{\qp,h,\q}$, represented by the lowest gray disk in Fig. \ref{fig:amplitude squared e+e-}, indicates the quantum amplitude for the splitting of a quark $\q$ in a hadron $h$ and a left-over quark $\qp$. It will be referred to as the splitting matrix. 
The matrix elements of the splitting matrix are indicated by $\langle \helQuarkp|T_{\qp,h,\q}|\helQuark\rangle$, where the index $\helQuarkp$ labels the spin state of the quark $\qp$. We have assumed for the moment the hadron $h$ to be spin-less ($a=a'=0$). The same definitions apply to the splitting matrix $T_{\qpbar,H,\qbar}$ (second upper gray disk in Fig. \ref{fig:amplitude squared e+e-}) that describes the splitting of the antiquark $\qbar$ in the hadron $H$ and the left-over antiquark $\qpbar$. The corresponding matrix elements are $\langle \helAntiQuarkp|T_{\qpbar,H,\qbar}|\helAntiQuark\rangle$, where $\helAntiQuarkp$ labels the spin state of $\qpbar$. The hadron $H$ is also taken to be spin-less, for the moment. The splitting matrices for the quark and antiquark, taken from the string+${}^3P_0$ model in Ref. \cite{Kerbizi:2021M20}, are described in detail in Sec. \ref{sec:hadronization model}.

The quantity $\eta(\qp)$ is the \textit{acceptance} matrix of $\qp$. If quarks were not confined, it could characterize a polarized detector for $\qp$. In reality it carries the spin information coming ``backwards in time" from the analysis of particles further produced in the fragmentation chain (for the definition and the use of the acceptance matrices see Ref. \cite{Artru:2008cp}). The associated matrix elements are $\langle \helQuarkp'|\eta(\qp)|\helQuarkp\rangle$. If there is no information coming from the future emissions of $\qp$, or equivalently if that information is integrated over, the acceptance matrix is the identity matrix $\langle \helQuarkp'|\eta(\qp)|\helQuarkp\rangle=\delta_{\helQuarkp'\helQuarkp}$. The acceptance matrix $\eta(\qpbar)$ for the antiquark $\qpbar$ has a similar meaning. More rigorously, the acceptance matrices $\eta(\qp)$ and $\eta(\qpbar)$ should be gathered in one $4\times4$ matrix $\eta(\qp,\qpbar)$, which takes into account the spin correlations transmitted by closing the quark line. In this model we neglect such correlations, and decompose $\eta(\qp,\qpbar)=\eta(\qp)\otimes 
\eta(\qpbar)$.

The cross section in Eq. (\ref{eq:A^2 e+e-}) can be further simplified to obtain the expression
\begin{equation}\label{eq:A^2 e+e- simplified}
\begin{aligned}
   & d\sigma(e^+\!\! \uparrow e^-\!\! \uparrow\rightarrow h\,H\,X)\propto\langle\helQuark,\helAntiQuark|\M|\helElec,\helPos\rangle\,\rho_{\helElec\helElec'}^{e^-}\,\rho_{\helPos\helPos'}^{e^+}\, \\ &\langle\helElec',\helPos'|\M^{\dagger}|\helQuark',\helAntiQuark'\rangle \times \langle \helQuark'|\eta(\q)|\helQuark\rangle\,\langle \helAntiQuark'|\eta(\qbar)|\helAntiQuark\rangle \\
    &\equiv |\M|^2\,\Trqq\left[\rho(\q,\qbar)\,\eta(\q)\otimes \eta(\qbar)\right].
\end{aligned}
\end{equation}
The different pieces of the cross section are gathered to define the spin-summed squared amplitude $|\M|^2$ associated to the hard subprocess, the joint spin density matrix $\rho(\q,\qbar)$ of the $\q\qbar$ pair (represented in Fig. \ref{fig:amplitude squared e+e-} by the rectangular domain A), and the acceptance matrices $\eta(\q)$ for the initial quark $\q$ and $\eta(\qbar)$ for the initial antiquark $\qbar$. The operation $\Trqq$ indicates the trace over the spin indices of $\q$ and $\qbar$.

\begin{figure}[tb]
\centering
\begin{minipage}[b]{0.5\textwidth}
\includegraphics[width=0.65\textwidth]{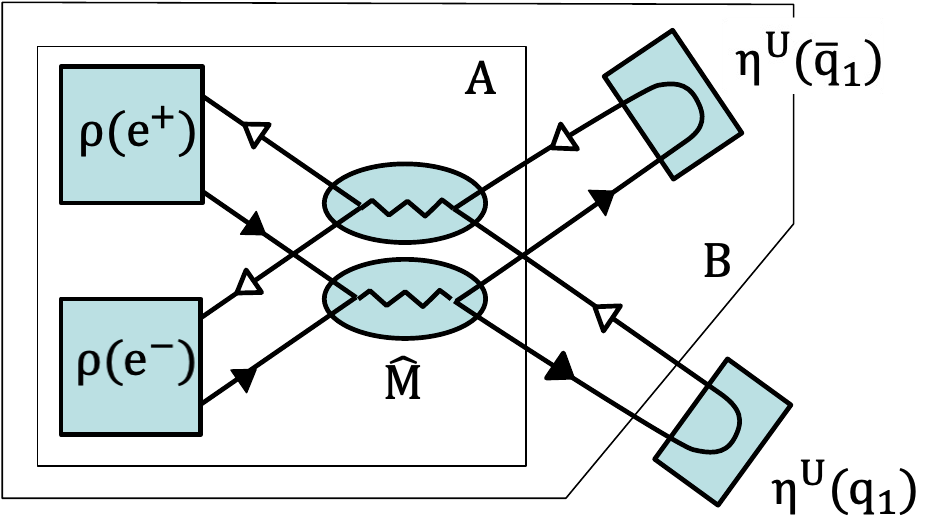}
\end{minipage}
\caption{Folded unitarity diagram for the reaction $e^+\!\!\uparrow e^-\!\!\uparrow \ \to \q\qbar$. $\eta^{\rm U}$ is the unit 2$\times$2 acceptance matrix. Rectangle A: $\la |{\it \hat M}|^2 \ra\rho(\q,\qbar)$. Domain B: $\la |{\it \hat M}|^2 \ra \rho(\q)$ (Eq. (\ref{eq:A^2 e+e- -> h X})).}
\label{fig:ee->qq}
\end{figure}

The squared amplitude $|\M|^2$ is related to the cross section of the hard scattering $e^+ \!\! \uparrow e^- \!\! \uparrow \rightarrow \q\qbar$ for not-analyzed quarks, and it is given by
\begin{equation}\label{eq:M^2}
    |\M|^2 \equiv \langle\helQuark,\helAntiQuark|\M|\helElec,\helPos\rangle\,\rho_{\helElec\helElec'}^{e^-}\,\rho_{\helPos,\helPos'}^{e^+}\, \langle\helElec',\helPos'|\M^{\dagger}|\helQuark,\helAntiQuark\rangle.
\end{equation}
It is represented by the folded unitarity diagram in Fig. \ref{fig:ee->qq}.

\begin{figure}[tbh]
\centering
\begin{minipage}[b]{0.5\textwidth}
\includegraphics[width=0.65\textwidth]{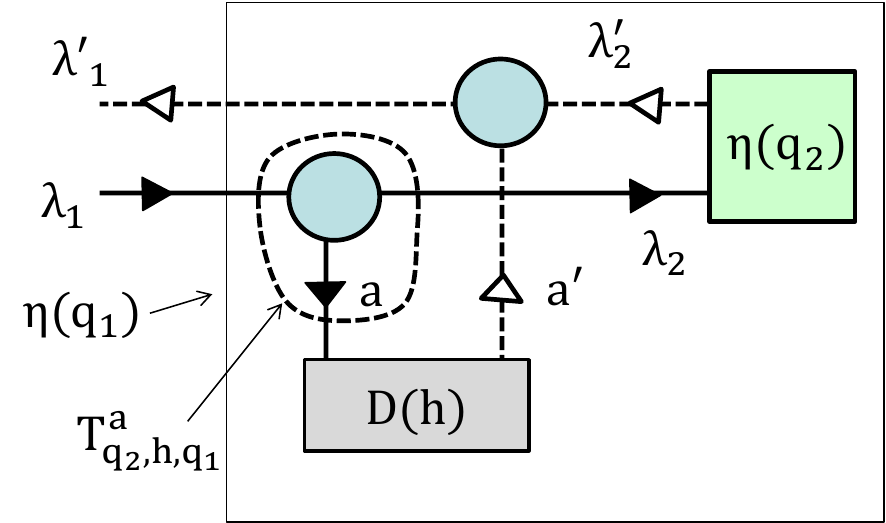}
\end{minipage}
\caption{Derivation of the acceptance matrix $\eta(\q)$ from $\eta(\qp)$ and $\eta(h) \equiv D(h)$. It applies to Eqs.(\ref{eq:acceptace q and qbar})-(\ref{eq:acceptace q and qbar matrix form}), taking $\eta(h)=1$.}
\label{fig:eta}
\end{figure}

Concerning the acceptance matrices $\eta(\q)$ and $\eta(\qbar)$, they can be obtained by comparing Eq. (\ref{eq:A^2 e+e- simplified}) with Eq. (\ref{eq:A^2 e+e-}). The expressions are
\begin{subequations}
\begin{equation}\label{eq:acceptace q and qbar}
\begin{aligned}
    \langle \helQuark'|\eta(\q)|\helQuark\rangle&=\langle \helQuark'|T^{\dagger}_{\qp,h,\q}|\helQuarkp'\rangle\, \langle \helQuarkp'|\eta(\qp)|\helQuark\rangle\, \langle \helQuark|T_{\qp,h,\q}|\helQuark\rangle, \\
    \langle \helAntiQuark'|\eta(\qbar)|\helAntiQuark\rangle&=\langle \helAntiQuark'|T^{\dagger}_{\qpbar,H,\qbar}|\helAntiQuarkp'\rangle\,\langle \helAntiQuarkp'|\eta(\qpbar)|\helAntiQuark\rangle\, \langle \helAntiQuark|T_{\qpbar,H,\qbar}|\helAntiQuark\rangle,
\end{aligned}
\end{equation}
and when written in matrix form they are
\begin{equation}\label{eq:acceptace q and qbar matrix form}
\begin{aligned}
    \eta(\q) &=T^{\dagger}_{\qp,h,\q}\, \eta(\qp)\,T_{\qp,h,\q} \\
    \eta(\qbar) &=T^{\dagger}_{\qpbar,H,\qbar}\,\eta(\qpbar)\, T_{\qpbar,H,\qbar}.
\end{aligned}
\end{equation}
\end{subequations}
These matrices bring to the hard scattering the spin information from the splittings of $\q$ and $\qbar$. The diagrammatic representation of $\eta(\q)$ is shown in Fig. \ref{fig:eta} for $h$ of arbitrary spin. For the spinless case, $a=a'=0$, $\eta(h)=1$. $\eta(\q)$ is also indicated in Fig. \ref{fig:amplitude squared e+e-} by the rectangular domain B. A similar diagram can be drawn for $\eta(\qbar)$.

From Eq. (\ref{eq:A^2 e+e- simplified}) one can obtain the cross section for single hadron production $e^+\!\!\uparrow e^-\!\!\uparrow\rightarrow h X$ by taking $\eta(\qbar)=\Iqbar$. It is
\begin{subequations}
\begin{equation}\label{eq:A^2 e+e- -> h X}
\begin{aligned}
   & d\sigma(e^+\!\!\uparrow e^-\!\!\uparrow \rightarrow h\,X)\propto |\M|^2\,\Tr\left[\rho(\q)\,\eta(\q)\right],
\end{aligned}
\end{equation}
where 
\begin{equation}\label{eq:rho(q1)}
    \rho(\q)=\Tr_{\qbar}\,\rho(\q,\qbar)
\end{equation}
\end{subequations}
is the spin density matrix of $\q$ obtained by the partial trace over the spin indices of $\qbar$. It is represented in Fig. \ref{fig:ee->qq} by the domain B. The cross section in Eq. (\ref{eq:A^2 e+e- -> h X}) is represented by the full diagram of Fig. \ref{fig:ee->qq}, but removing the U-turn in the lower-right gray rectangle and the upper index U for $\eta(\q)$.
It is well known that the single hadron production in $e^+e^-$ can not be used for studying transverse spin effects due to the fact that the quark (and the antiquark) is unpolarized, as can be seen from Eq. (\ref{eq:rho(q1)}) and Eq. (\ref{eq:initial C coefficients}).


\subsection{The hard scattering}\label{sec:kinematics}
\begin{figure}[b]
\centering
\begin{minipage}[b]{0.48\textwidth}
\includegraphics[width=0.95\textwidth]{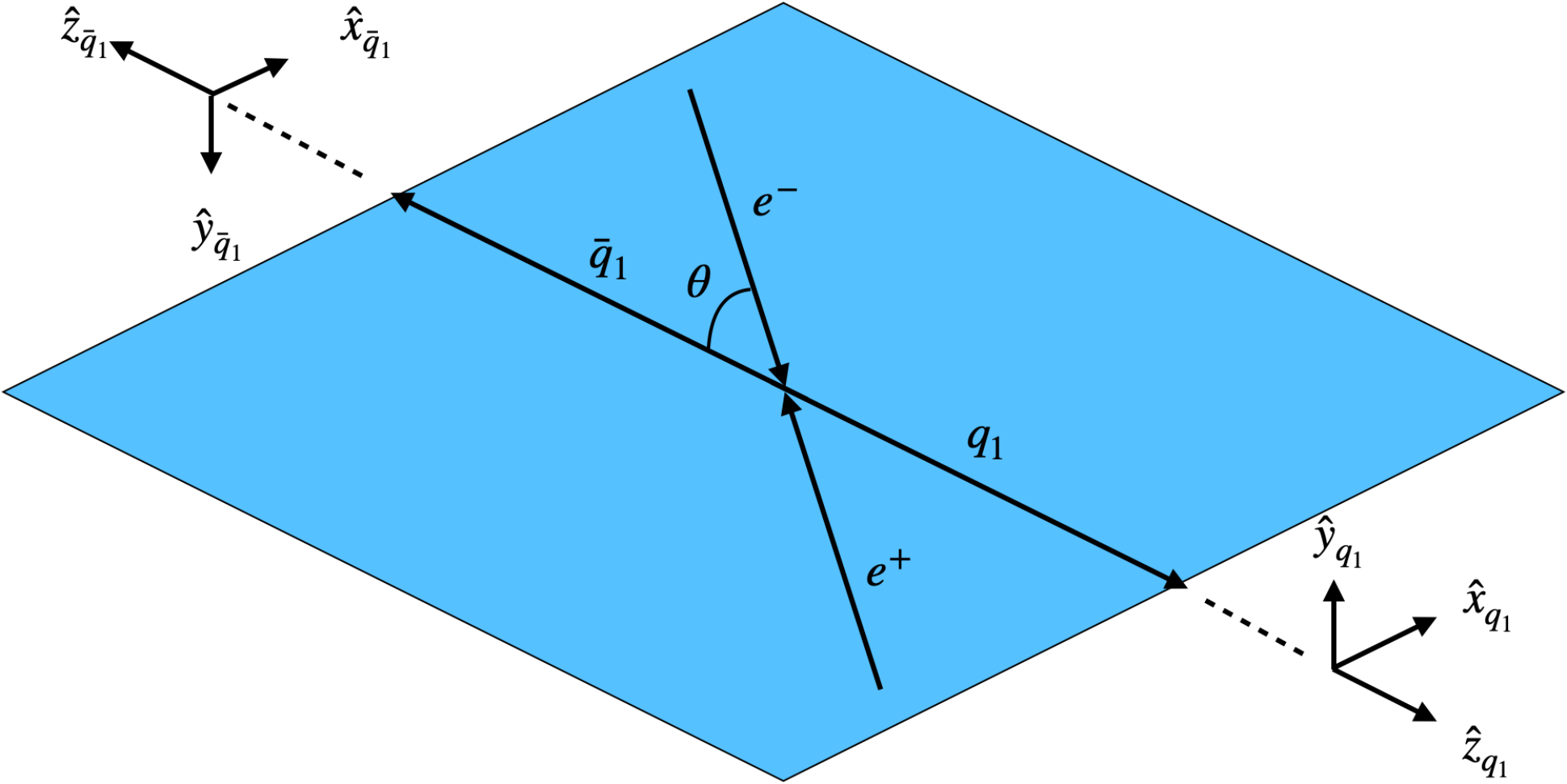}
\end{minipage}
\caption{Kinematics of the 
annihilation process $e^+e^-\rightarrow \q\qbar$ in the c.m.s. Also shown are the axes of the helicity frames of the quark and the antiquark.}
\label{fig:e+e- process}
\end{figure}

The kinematics of the hard scattering $e^+(p_+)\,e^-(p_-)\rightarrow \q(\k)\,\qbar(\kbar)$, where in the parenthesis we have indicated the four-momentum of each particle, is shown in Fig. \ref{fig:e+e- process} in the center of mass system (c.m.s). As already mentioned in the introduction, we neglect the gluon radiation from the quarks. 
The four-momenta are given by
\begin{align}\label{eq:momenta}
\nonumber    p_{\mp} &= \frac{\sqrt{s}}{2}\,(1,\mp\sin\theta,0,\pm\cos\theta), \\
     \k &= \frac{\sqrt{s}}{2} \, (1,0,0, \betaq), && \kbar = \frac{\sqrt{s}}{2} \, (1,0,0, -\betaq). 
\end{align}
The electron mass is neglected as it is much smaller than the considered values of the center of mass energy $\sqrt{s}=\sqrt{2\,p_+ \cdot p_-}$. $\theta$ is the angle between the $\textbf{p}_-$ and $\kQuark$.
The quantity $\betaq=\sqrt{1-4\,m_q^2/s}$ is the velocity of the quark and $m_q$ is the quark mass\footnote{The quark mass is not neglected to account for the mass of charm quarks that is not negligible when compared to the center of mass energy of the BESIII experiment. Taking the mass of the charm quark to be $m_c\simeq 1.5\,\GeV/c^2$, the velocity of charm quarks in the c.m.s at the BESIII energy is expected to be $\betaq\simeq 0.55$.}. 

The hard scattering differential cross section is related to the squared matrix element in Eq. (\ref{eq:M^2}) by $d\hat{\sigma}/d\cos\theta=3\beta_q\,|\M|^2/(32\pi s)$, where the factor $3$ is included to account for the number of quark colors. Using the expressions for the helicity amplitudes in Appendix \ref{sec:helicity amplitudes}, we obtain the known angular distribution \cite{ParticleDataGroup:2022pth}
\begin{equation}\label{eq:hard cross section}
    \begin{aligned}
        \frac{d\hat{\sigma}(\q\qbar)}{d\cos\theta} =  \frac{3\pi\alpha^2}{2s}\,e_q^2\,\beta_q\,[1+\cos^2\theta+(1-\beta_q)^2\sin^2\theta].
    \end{aligned}
\end{equation}

In the c.m.s we introduce the so-called helicity frames of the quark and the antiquark. They are the reference systems attached to the quark and the antiquark that are introduced when calculating the helicity amplitudes in Appendix \ref{sec:helicity amplitudes} \footnote{The definition of the helicity frame can be found, e.g., in Ref. \cite{Leader:2011vwq}. See also Ref. \cite{DAlesio:2021dcx}.}. We represent the quark helicity frame (QHF) with the set of axes $\lbrace \xq, \yq, \zq \rbrace$, and the antiquark helicity frame (AHF) with the set of axes $\lbrace \xqbar, \yqbar, \zqbar\rbrace$. The axes are defined to be
\begin{equation}\label{eq:QuarkHelicityFrame}
\begin{aligned}
    \zq&=\frac{\kQuark}{|\kQuark|}, && \yq = \frac{\pElec \times \zq}{|\pElec \times \zq|}, && \xq = \yq \times \zq, \\
    \zqbar&=\frac{\kAntiQuark}{|\kAntiQuark|}, && \yqbar = \frac{\pElec \times \zqbar}{|\pElec \times \zqbar|}, && \xqbar = \yqbar \times \zqbar.
\end{aligned}
\end{equation}
The helicity frames of the quark and the antiquark are shown in Fig. \ref{fig:e+e- process}. The two frames share their $x$-axes, whereas the other axes are the opposite. The four-momenta given in Eq. (\ref{eq:momenta}) are expressed in the QHF.

Considering a generic vector $\textbf{v}$ with azimuthal angle $\phi^{(\QHF)}(\textbf{v})$ and polar angle $\theta^{(\QHF)}(\textbf{v})$ measured in the QHF, and azimuthal angle $\phi^{(\AHF)}(\textbf{v})$ and polar angle $\theta^{(\AHF)}(\textbf{v})$ measured in the AHF, the relations among the angles expressed in the two frames are
\begin{align}\label{eq:angles}
    \phi^{(\AHF)}(\textbf{v}) = 2\pi - \phi^{(\QHF)}(\textbf{v}), && \theta^{(\AHF)}(\textbf{v}) = \pi - \theta^{(\QHF)}(\textbf{v}).
\end{align}
These relations are useful to express observables in the same reference system, and will be used in Sec. \ref{sec:application}.

\subsection{The joint spin-density matrix of the $\q\qbar$ pair}\label{sec:joint spin density matrix}
In the annihilation $e^+e^-\rightarrow \gamma^*\rightarrow \q\qbar$, the virtual photon has spin one and has a non-zero tensor polarization. 
In addition to a specific $\theta$ dependence, this induces correlations among the spin states of $\q$ and $\qbar$, which can be encoded in the joint spin density matrix $\rho(\q,\qbar)$ of the $\q\qbar$ system.

The expression for the density matrix $\rho(\q,\qbar)$ can be obtained by comparing the third and the first lines of Eq. (\ref{eq:A^2 e+e- simplified}), and it is
\begin{equation}\label{eq:joint density matrix}
\begin{aligned}
\rho(\q,\qbar) &=(|\M|^2)^{-1}\,\M_{\helElec\helPos;\helQuark\helAntiQuark}\,\rho^{e^-}_{\helElec\helElec'}\,\rho^{e^+}_{\helPos\helPos'}\,\M^{*}_{\helElec'\helPos';\helQuark'\helAntiQuark'}\,\,\\ &\times |\helQuark',\helAntiQuark'\rangle\langle \helQuark,\helAntiQuark| \equiv \frac{1}{4}\,\Cqq_{\alpha\beta}\,\sigmaq_{\alpha}\otimes \sigmaqbar_{\beta}.
\end{aligned}
\end{equation}
In the second equality the density matrix is decomposed along a basis spanned by the tensor product of the Pauli matrices $\sigmaq_{\alpha}=(\Iq,\sigmaXq,\sigmaYq,\sigmaZq)$ for the quark and $\sigmaqbar_{\beta}=(\Iqbar,\sigmaXqbar,\sigmaYqbar,\sigmaZqbar)$ for the antiquark. The indices $\alpha$ and $\beta$ take the values $0,x,y,z$.
The matrix $\Iq$ ($\Iqbar$) indicates the identity matrix in the spin space of $\q$ ($\qbar$). The superscript $\q$ ($\qbar$) indicates the projection of the vector of Pauli matrices $\boldsymbol{\sigma}=(\sigma^x,\sigma^y,\sigma^z)$ along the axes of the QHF (AHF).
The coefficients $\Cqq_{\alpha\beta}$ introduced in the second line in Eq. (\ref{eq:joint density matrix}) express the correlations between the spin states of $\q$ and $\qbar$, and will be referred to as the correlation coefficients. They can be obtained by taking the trace of Eq. (\ref{eq:joint density matrix}) with $\sigmaq_{\alpha}\otimes \sigmaqbar_{\beta}$. The joint spin density matrix is normalized such that $\Cqq_{00}=1$, and the factor $1/4$ assures that $\rm{Tr}_{\q,\qbar}\,\rho(\q,\qbar) = 1$.

The non-vanishing correlation coefficients can be calculated using the first line in Eq. (\ref{eq:joint density matrix}) and the helicity amplitudes in Tab. \ref{tab:helicity amplitudes} (see Appendix \ref{sec:helicity amplitudes}). We obtain for the decomposed density matrix
\begin{equation}\label{eq:initial C coefficients}
\begin{aligned}
& \rho(\q,\qbar) = \frac{1}{4}\,\bigg[\Iq\otimes \Iqbar - \frac{2-(2-\betaq^2)\,\sin^2\theta}{2-\betaq^2\sin^2\theta} \,\sigmaZq\otimes \sigmaZqbar \\
&+ \frac{(2-\betaq^2)\,\sin^2\theta}{2-\betaq^2\sin^2\theta}\, \sigmaXq\otimes \sigmaXqbar + \frac{\betaq^2\,\sin^2\theta}{2-\betaq^2\sin^2\theta}\,\sigmaYq\otimes \sigmaYqbar \\
&+ \frac{(1-\betaq^2)^{1/2}\,\sin 2\theta}{2-\betaq^2\sin^2\theta} \,(\sigmaXq\otimes \sigmaZqbar - \sigmaZq\otimes \sigmaXqbar)\bigg].
\end{aligned}
\end{equation}
This expression takes into account the non-vanishing quark mass through the dependence on the quark velocity $\betaq$. If the terms depending on the quark mass are neglected, the expression in Eq. (\ref{eq:initial C coefficients}) is the same as in Ref. \cite{Chen:1994ar} when only the terms arising from the exchange of the $\gamma^*$ are considered.


The joint density matrix in Eq. (\ref{eq:initial C coefficients}) is not separable, meaning that it can not be expressed as $\rho(\q,\qbar)=\sum_i\,w_i\, \rho_i(\q)\otimes \rho_i(\qbar)$ with positive weights $w_i$ and density matrices $\rho_i(\q)$ for the quark and $\rho_i(\qbar)$ for the antiquark. This is the general case that holds also for other processes, e.g. the $Z^0$ boson
decay $Z^0\rightarrow \q\antiq$ or the Higgs boson decay $H^0\rightarrow \q\antiq$. The case of a separable denstiy matrix, e.g. as in the decay $W^{\pm}\rightarrow q_1\bar{q}_1$ in the massless quark limit owing to the fact that the $W^{\pm}$ bosons couple left-handed particles, can be considered as one exception. 



\begin{figure}[tbh]
\centering
\begin{minipage}[b]{0.48\textwidth}
\includegraphics[width=1.0\textwidth]{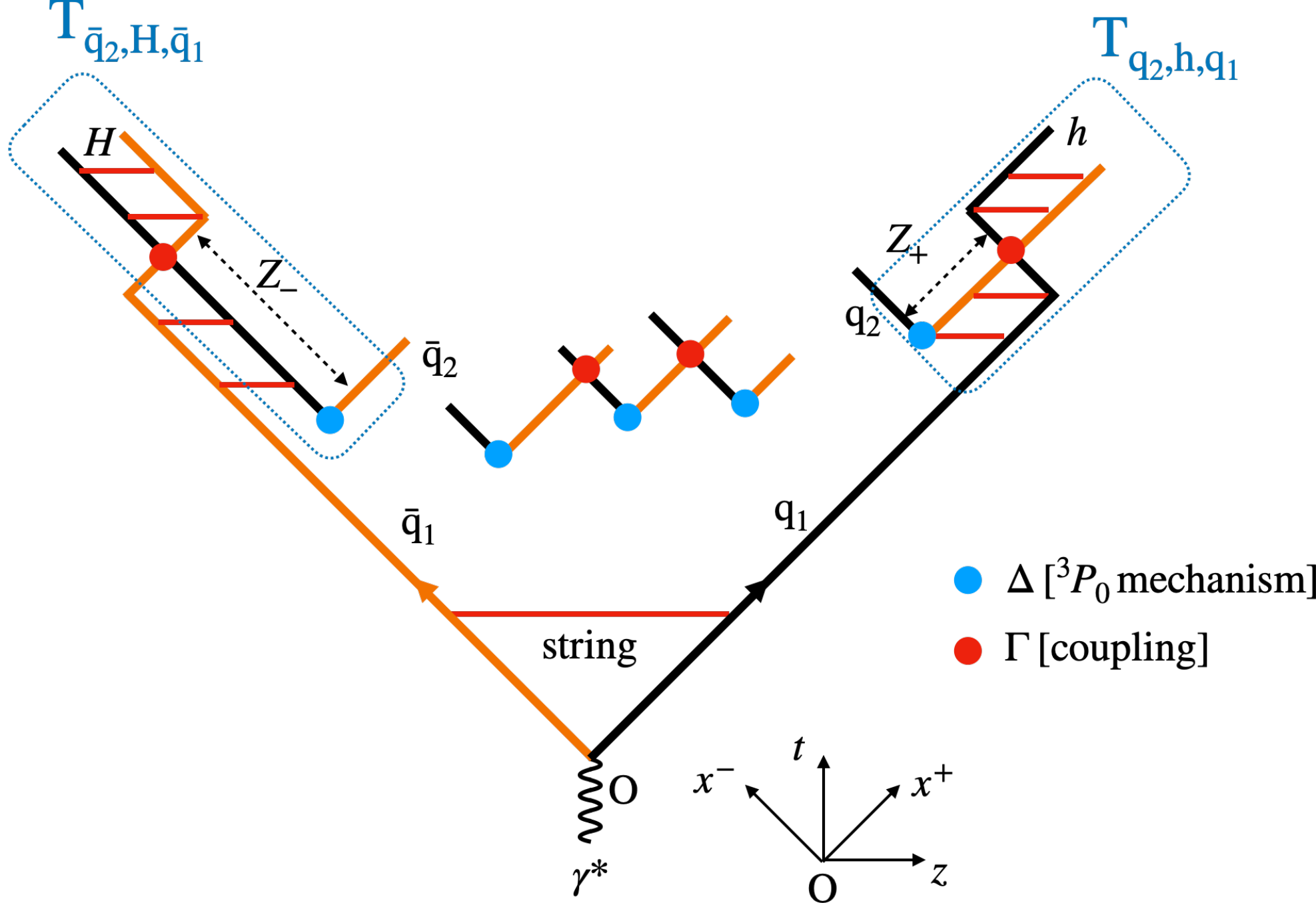}
\end{minipage}
\caption{Spacetime picture of the string fragmentation of the $\q\qbar$ pair. Also shown are the spacetime points where the string breaks via a quark-antiquark pair in the relative ${}^3P_0$ state and where quarks couple to the hadrons.}
\label{fig:string history}
\end{figure}

\subsection{The splitting matrix of the string+${}^3P_0$ model}\label{sec:hadronization model}
We describe the hadronization $\q\antiq\rightarrow h_1,h_2,\dots,h_N$ in the string fragmentation framework by using the string+${}^3P_0$ model in Ref. \cite{Kerbizi:2021M20}. The space-time picture of the hadronization of the $\q\qbar$ pair in the c.m.s. is depicted in Fig. \ref{fig:string history}. After being produced at the spacetime point indicated by $O$, the quark and the antiquark propagate in opposite directions stretching a straight string between them. The string axis is the quark-antiquark axis in the c.m.s, and we take the $\zu$ axis along the quark momentum. In the string fragmentation model, the hadronization can be viewed as the iteration of the elementary quark splitting $\q\rightarrow \h + \qp$ (top right part of Fig. \ref{fig:string history}), where the emitted hadron $\h=(\q\qpbar)$ can be either a PS meson or a VM (baryons and heavier hadronic states are not yet included in the string+${}^3P_0$ model). The leftover quark $\qp$ propagates the spin-information to the next splitting $q_2\rightarrow h'+q_3$ etc. We indicate the four-momenta of $\q$, $h$ and $\qp$ by $\k$, $\p$ and $\kp$, respectively. Momentum conservation implies $\p = \k - \kp$.  

The four-momentum of $\h$ is parametrized by introducing the longitudinal splitting variable $Z_+=\p^+/\k^+$, the transverse momentum $\pt$ and the transverse energy $\eh^2=\sqrt{\mh^2+\ptpt}$, $M$ being the hadron mass. The lightcone components for a generic four-vector $v$ are defined as $v^{\pm} = v^0\pm v^{z}$. The transverse momentum is $\pt=(p^x,p^y)$ and we have $\pt=\kt-\kpt$. 

The model describes the elementary splitting in momentum and spin-space by the means of the splitting matrix \cite{Kerbizi:2021M20} 
\begin{widetext}
\begin{align}\label{eq:T quark}
T_{\qp,\h,\q}(\mh^2,Z_+,\pt;\kt) = C_{\qp,\h,\q}\,D_{\h}(\mh^2)\,\left(\frac{1-Z_+}{\eh^2}\right)^{a/2}\, e^{-\bL\eh^2/(2Z_+)}\,N_a^{-1/2}(\eh^2)\,
\fT(\kptkpt)\, \Delta(\kpt)\, \Gammaq_{\h,\sh}\,\hat{u}^{-1/2}_{\q}(\kt).
\end{align}
\end{widetext}
The coefficient $C_{\qp,\h,\q}$ describes the splitting in flavor-space and it is based on the wave function of $\h$ in isospin space.
The function $|D_{\h}(\mh^2)|^2$ gives the invariant mass distribution of $\h$. For a fixed meson mass it is a delta function centered on the nominal squared mass $m_h^2$, while for a resonance it is a relativistic Breit-Wigner function with mass and width fixed to their nominal values (more details can be found in Ref. \cite{Kerbizi:2021M20}).
The factors involving the $Z_+$ variable describe the distribution of the longitudinal momentum of $\h$, which depends on the free parameter $\bL$. The function $N_a(\eh^2)$ depends on the squared transverse energy of $\h$ and plays the role of a normalization factor for the $Z_+$-dependent part of the splitting amplitude squared. It is given by $N_a^{-1}(\eh^2)=\int_0^1\,dZ_+\,Z_+^{-1}\,\left[(1-Z_+)/\eh^2\right]^a\,\exp(-\bL\eh^2/Z_+)$.
The function $\fT$ provides the transverse momentum cutoff for the quarks created at string breaking and it is taken to have the exponential form $\fT(\kptkpt)=\exp(-\bT\,\kptkpt/2)$, with $\bT$ being a free parameter. 

The last three terms of Eq. (\ref{eq:T quark}) express the spin-dependence of the splitting amplitude. The matrix $\Delta(\kpt)=\mu+\i\boldsymbol{\sigma}\cdot(\zq\times\kpt)=\mu+\i\sigmaZq\boldsymbol{\sigma}\cdot\kpt$ parametrizes the ${}^3P_0$ wave function of the $\qp\qpbar$ pair produced at the string breaking (see Fig. \ref{fig:string history}) and it depends on the complex mass $\mu$. The latter can in principle depend on the flavor of $\qp$ and on $\kptkpt$, but it is taken to be flavor and $\kptkpt$ independent. The imaginary part $\rm{Im}(\mu)$ is responsible for transverse spin effects, e.g. the Collins effect and the dihadron production asymmetry, while $\rm{Im}(\mu^2)=2\,\rm{Re}(\mu)\,\rm{Im}(\mu)$ is responsible for longitudinal spin effects, e.g. the jet-handedness \cite{Kerbizi:2018qpp}. The matrix $\sigmaqT=(\sigma_x, \sigma_y)$ is the vector of Pauli matrices with only transverse components.

The matrix $\Gammaq_{\h,\sh}$ projects the spin state of the $\q\qpbar$ pair onto the spin state of the hadron $\h$ (see Fig. \ref{fig:string history}), and it is referred to as the coupling matrix. It is given by
\begin{equation}\label{eq:Gamma q}
   \Gammaq_{\h,\sh} =  \begin{cases}
   \sigma_{z}  & \rm{if\, \h = PS }\\ 
    \GL\,\Iden\,\VL^* + \GT\,\sigmaqT\sigma_{z}\cdot\VT^* & \rm{if\, \h = VM}                 
   \end{cases}, 
\end{equation}
where $\V =(\VT,\VL)$ is the linear polarization vector of the VM. In the case of VM emission, the coupling matrix depends on the complex coupling constants $\GL$ and $\GT$, which describe the coupling of $\q$ and $\qp$ with a VM having longitudinal and transverse polarization with respect to the string axis, respectively. Only the combinations $\fVM = 2|\GT|^2+|\GL|^2|$, $\fL=|\GL|^2/(2|\GT|^2+|\GL|^2)$ and $\thetaLT=\arg(\GL/\GT)$ are, however, relevant. The parameter $\fVM$ gives the ratio between the probability of producing a VM and the probability of producing a PS meson in the elementary splitting. The parameters $\fL$ and $\thetaLT$ govern the fraction of longitudinally polarized VMs in each splitting and their oblique polarization, respectively.

In the present model, due to the particular choice for $N_a(\varepsilon_h^2)$ \cite{Kerbizi:2019ubp}, the function $\hat{u}_{\q}(\kt)$ does not depend neither on $\kt$ nor on the polarization state of $\q$ \cite{Kerbizi:2021M20}. The latter can be decomposed as
\begin{align}\label{eq:u matrix}
\nonumber   \hat{u}_{\q} =& \sum_h\displaystyle\,\hat{u}_{\q,h}, \\
\nonumber   \hat{u}_{\q,\h} =& |C_{\qp,\h,\q}|^2\,(|\mu|^2+\langle \kTkT\rangle_{\fT})\, \\
   & \times \begin{cases}
   1  & \rm{if\, h = PS }\\ 
   \fVM & \rm{if\, h = VM}                 
   \end{cases}, 
\end{align}
where we have defined $\langle \kTkT\rangle_{\fT} = \int\,d^2\kT\,\kTkT\,\fT^2(\kTkT)/\int\,d^2\kT\,\fT^2(\kTkT)$. The ratio $\hat{u}_{\q,\h}/\hat{u}_{\q}$ gives the relative probability of producing the hadron species $\h$ in the elementary splitting.

Due to the left-right (LR) symmetry \cite{Andersson:1983ia} the string fragmentation process can be viewed equivalently as the iteration of the elementary splitting $\qbar\rightarrow \hbar +\qpbar$ of the antiquark $\qbar$ in the emitted hadron $H=(\qbar \qp)$ and the leftover antiquark $\qpbar$ (top left part of Fig. \ref{fig:string history}). To describe the elementary splitting of an antiquark we indicate by $\kbar$ and $\kpbar$ the four-momenta of $\qbar$ and $\qpbar$, and by $P$ the four-momentum of $H$. The transverse momenta of $\qbar$, $H$ and $\qpbar$ are indicated by $\ktbar$, $\Pt$ and $\kptbar$, respectively. Momentum conservation in the spliting implies $\Pt =\ktbar-\kptbar$.

The splitting matrix for the antiquark splitting was not given in Ref. \cite{Kerbizi:2021M20}. It can however be obtained from Eq. (\ref{eq:T quark}) using the LR symmetry, and it is
\begin{widetext}
\begin{align}\label{eq:T antiquark}
T_{\qpbar,\hbar,\qbar}(\mh^2,Z_-,\Pt;\ktbar) = C_{\qpbar,\hbar,\qbar}\,D_{\hbar}(\mh^2)\,\left(\frac{1-Z_-}{\ehbar^2}\right)^{a/2}\,e^{-\bL\ehbar^2/(2Z_-)}\,N_a^{-1/2}(\ehbar^2) \, \fT(\kptbarkptbar) \, \Delta(\kptbar)\,\Gammaqbar_{H,s_H}\,\hat{u}^{-1/2}_{\qbar}(\ktbar),
\end{align}
\end{widetext}
where it is $\Delta(\kptbar)=\mu+\i\boldsymbol{\sigma}\cdot(\zqbar\times\kptbar)=\mu+\i\sigmaZqbar\boldsymbol{\sigma}\cdot\kptbar$. It is the same expression as Eq. (\ref{eq:T quark}) with the substitutions $\q\rightarrow \qbar$, $h\rightarrow H$, $\qp\rightarrow \qpbar$ and $Z_+\rightarrow Z_-$, and $\lbrace\kt,\pt,\kpt\rbrace \rightarrow \lbrace \ktbar,\Pt,\kptbar\rbrace$. The variable $Z_-$ is defined as $Z_-=P^-/\kbar^-$ (see Fig. \ref{fig:string history}).

\section{The polarized fragmentation of a string with entangled quarks}\label{sec:polarized string fragmentation}
The string+${}^3P_0$ model has been applied to the fragmentation of a string stretched between a quark $\q$ on the one endpoint and an antiquark $\qbar$ (or a diquark) on the other endpoint, where only the quark polarization is considered \cite{Kerbizi:2018qpp,Kerbizi:2019ubp,Kerbizi:2021M20}. The fragmentation chain is thus developed from the quark towards the other endpoint, and the spin information is propagated only from the quark side.

In this section we extend the string+${}^3P_0$ model of Ref. \cite{Kerbizi:2021M20} to the fragmentation of a string stretched between $\q$ and $\qbar$ with correlated spin states as described by the joint spin density matrix $\rho(\q,\qbar)$. This development is needed for the Monte Carlo simulation of $e^+e^-$ annihilation to hadrons including the quark spin effects. The final recursive recipe for the simulation of $e^+e^-$ annihilation is given in Sec. \ref{sec:recursive recipe}. 


\subsection{The steps of the fragmentation chain}\label{sec:steps}
\subsubsection{Hadron emission from the quark end}\label{sec:hadron emission quark}
\begin{figure}[tb]
\centering
\begin{minipage}[b]{0.5\textwidth}
\includegraphics[width=0.549\textwidth]{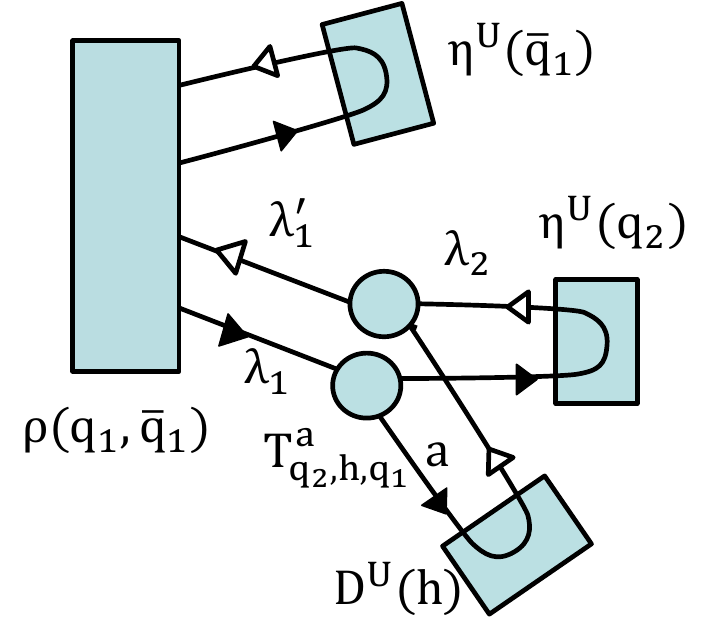}
\end{minipage}
\caption{Diagrammatic representation of the splitting functions in Eq. (\ref{eq:splitting F_PS}) and in Eq. (\ref{eq:splitting F_VM}).}
\label{fig:splitting F q}
\end{figure}

\paragraph{Pseudoscalar meson emission.}
We start by considering the emission of a PS meson $\h$ from the quark end of the string, i.e. by the splitting $\q\rightarrow h +\qp$. The emission of $h$ is described by the probability distribution of emitting the hadron with a given four-momentum. 
This is obtained neglecting the information coming from the $\qbar$ end. Inserting in Eq. (\ref{eq:A^2 e+e- simplified}) the expression of $\eta(\q)$ given in Eq. (\ref{eq:acceptace q and qbar}), and imposing $\eta(\qbar)=\Iqbar$ and $\eta(\qp)=\Iqp$\footnote{This assumption bears on the fact that, for large enough invariant mass of the remaining $\qp\qbar$ system, the spin information decays along the quark fragmentation chain of this system (see Eqs. (31)-(32) 
of Ref. \cite{Kerbizi:2019ubp}).}, we obtain the probability distribution
\begin{align}\label{eq:splitting F_PS}
\nonumber  & \frac{dP(\q\rightarrow h=\PS + \qp; \q\qbar)}{dZ_+Z_+^{-1}\,d^2\pt} \equiv F_{\qp,h,\q}(Z_+,\pt;\kt,\C^{\q\qbar}) \\
&=\rm{Tr}_{\qp\qbar}\left[\T_{\qp,h,\q}\,\rho(\q,\qbar)\,\T_{\qp,h,\q}^{\dagger} \right],
\end{align}
where we have defined $\textbf{T}_{\qp,h,\q}=T_{\qp,h,\q}\otimes \Iqbar$. The probability distribution is thus obtained by acting with the splitting matrix $T_{\qp,h,\q}$ on the quark spin sub-space of the joint spin density matrix and with the identity matrix $\Iqbar$ on the antiquark spin sub-space (there is yet no information about the emissions from the $\qbar$ end), and by taking the trace over the quark and antiquark spin indices. 
This is represented by the diagram in Fig. \ref{fig:splitting F q}, ignoring here the buckle to $D^{\rm U}(h)$. The differential probability $dP(\q\rightarrow h=\PS + \qp)$ in the first line in Eq. (\ref{eq:splitting F_PS}) is divided by the phase space element $dZ_+\,Z_+^{-1}\,d^2\pt$.

The function $F_{\qp,h,\q}(Z_+,\pt;\kt,\Cqq)$ will be referred to as the splitting function. It describes the energy-momentum sharing between $\h$ and $\qp$; it depends on the splitting variable $Z_+$ and on the transverse momentum $\pt$, given the value of the transverse momentum $\kt$ of $\q$ and the values of the correlation coefficients $\Cqq$ that implement the spin-correlations between the two string endpoints. The explicit expression of the splitting function for PS meson emission is obtained inserting in Eq. (\ref{eq:splitting F_PS}) the Eqs. (\ref{eq:T quark})-(\ref{eq:u matrix}). The result is (here it depends on $\kt=\textbf{0}$, but for iterations of the splitting we will have $\textbf{k}_{\rm n\,T}\neq \textbf{0}$)
\begin{align}\label{eq:splitting F_PS explicit}
\nonumber    F&_{\qp,h=\PS,\q}(Z_+,\pt;\kt,\C^{\q\qbar})
= \frac{\hat{u}_{\q,h}}{\hat{u}_{\q}} \left(\frac{1-Z_+}{\eh^2}\right)^{a} e^{-\bL\eh^2/Z_+}\, \\
\nonumber &\times N_a^{-1}(\eh^2)\, \fT^2(\kptkpt)\,\frac{|\mu|^2+\kptkpt}{|\mu|^2+\langle \kTkT\rangle_{\fT}}\\
&\times \left(1+\hat{a}(\kptabs)\,\Cqq_{x0}\,\sin\phikptQ-\hat{\mathnormal{a}}(\kptabs)\,\Cqq_{y0}\,\cos\phikptQ\right).
\end{align}
The first line describes the splitting in flavor space and the distribution of the $Z_+$ variable. The second line gives the distribution of the modulus of the transverse momentum $\kpt$ of $\qp$, while the third line gives the distribution of the azimuthal angle $\phikptQ$ of $\kpt$. The azimuthal angle $\phikptQ$ is measured in the QHF. The azimuthal distribution of $\textbf{k}_{2}$ has a modulation whose amplitude depends on the correlation coefficients of the string endpoints times the analysing power $\hat{a}$ (see Eq. (\ref{eq:splitting F_PS explicit})). The latter is given by
\begin{equation}
    \hat{a}(\kptabs) = \frac{2\rm{Im}(\mu)\,\kptabs}{|\mu|^2+\kptkpt}.
\end{equation}

To recall the meaning of $\hat{a}$, we consider the fragmentation of a quark with polarization vector $\Sq=(\SqT,\SqL)$, where $\SqT$ and $\SqL$ are the transverse and longitudinal polarizations with respect to the string axis. As shown in Ref. \cite{Kerbizi:2021M20}, the splitting function for the emission of a PS meson has a similar expression as Eq. (\ref{eq:splitting F_PS explicit}) but the third line is substituted with $1-\hat{a}\,\SqT\cdot\kptilde$. The distribution of the azimuthal angle of $\kpt$ is characterized by the modulation $\sin(\phikptQ-\phiSq)$, which is transferred to the hadron and is responsible for the Collins effect. 
Comparing with Eq. (\ref{eq:splitting F_PS explicit}), one can see that the coefficients $\Cqq_{x0}$ and $\Cqq_{y0}$ play the role of the $x$ and $y$ components of the transverse polarization vector $\SqT$. Note however that due to Eq. (\ref{eq:initial C coefficients}) these coefficients are zero for the primary quark-antiquark pair $\q\qbar$, so the last line of Eq. (\ref{eq:splitting F_PS explicit}) is just equal to unity. This will be no more the case in the iterations, i.e. after the first quark or antiquark splitting.

\paragraph{Vector meson emission.}
If the quark end of the string emits a VM instead of a PS meson, the acceptance matrix $\eta(\q)$ is more explicitly represented by the diagram of Fig. \ref{fig:eta} than by the rectangle B in Fig. \ref{fig:amplitude squared e+e-}. This diagram represents the probability of the process $\q\rightarrow h=\VM +\qp$. The splitting matrix in Eq. (\ref{eq:T quark}) can be decomposed as $T^{a}_{\qp,h=\VM,\q}\,\V_a$ and the VM coupling in Eq. (\ref{eq:Gamma q}) as $\Gamma^a\V_a$, where the label $a=x,y,z$ indicates the components of the polarization vector $\V$ of the VM in the QHF. 


Following Fig. \ref{fig:eta}, the acceptance matrix $\eta(\q)$ after the emission of a VM can be written as
\begin{equation}\label{eq:eta(q) VM}
\eta(\q)=T^{\dagger\,a'}_{\qp,h=\VM,\q}\,\eta(\qp)\,T^{a}_{\qp,h=\VM,\q}\,D_{a',a},
\end{equation}
where the matrix $D$ with elements $D_{a'a}=\langle a'|D|a\rangle$ is the decay matrix that implements the information about the decay products of the VM (see Sec. \ref{sec:decays}). The decay matrix is an acceptance matrix that replaces the trivial one $\eta(h)=1$ of the PS case. If the decay of the meson is not analyzed or the angular distribution of the decay products is integrated over, the decay matrix is the identity matrix $D_{a'a}=\delta_{a'a}$.

Inserting Eq. (\ref{eq:eta(q) VM}) in Eq. (\ref{eq:A^2 e+e- simplified}), the probability distribution for the emission of a VM from the quark end of the string is obtained by taking $\eta(\qbar)=\Iqbar$ and $\eta(\qp)=\Iqp$. 
This leads to the probability distribution for a non-analyzed VM
\begin{align}\label{eq:splitting F_VM}
\nonumber    &\frac{dP(\q\rightarrow h=\VM + \qp; \q\qbar)}{d\mh^2\,dZ_+Z_+^{-1}\,d^2\pt} \equiv F_{\qp,h,\q}(\mh^2,Z_+,\pt;\kt,\C^{\q\qbar}) \\
&= \rm{Tr}_{\qp\qbar}\left[\T^{a}_{\qp,h=\VM,\q}\,\rho(\q,\qbar)\,\T^{\dagger\,a}_{\qp,h=\VM,\q} \right],
\end{align}
where we have defined $\T^a_{\qp,h=\VM,\q}=T^{a}_{\qp,h=\VM,\q}\otimes\Iqbar$. Equation (\ref{eq:splitting F_VM}) is represented by the diagram in Fig. \ref{fig:splitting F q}. Compared to the PS case in Eq. (\ref{eq:splitting F_PS explicit}), the splitting function for VM emission 
depends additionally on the invariant mass squared $\mh^2$ of the meson, which is not fixed. The invariant mass is included also in the phase space factor to account for the mass distribution of the resonance.
As indicated in the second line in Eq. (\ref{eq:splitting F_VM}) a summation over the polarization states of the VM is understood.

Inserting in Eq. (\ref{eq:splitting F_VM}) the splitting matrix (\ref{eq:T quark}) with the VM coupling (\ref{eq:Gamma q}) and $\hat{u}_{\q}$ from (\ref{eq:u matrix}), we obtain the following expression for the splitting function for VM emission
\begin{align}\label{eq:splitting F_VM explicit}
\nonumber    F&_{\qp,h=\VM,\q}(\mh^2,Z_+,\pt;\kt,\C^{\q\qbar})
= \frac{\hat{u}_{\q,h}}{\hat{u}_{\q}}\, |D_{\h}(\mh^2)|^2\, \\
\nonumber &\times \left(\frac{1-Z_+}{\eh^2}\right)^{a}\,e^{-\bL\eh^2/Z_+}\, N_a^{-1}(\eh^2)\, \\
\nonumber &\times \fT^2(\kptkpt)\,\frac{|\mu|^2+\kptkpt}{|\mu|^2+\langle \kTkT\rangle_{\fT}}\\
\nonumber &\times \left(1-\hat{a}(\ktabs)\,\fL\,\Cqq_{x0}\,\sin\phikptQ+\hat{\mathnormal{a}}(\ktabs)\,\fL\,\Cqq_{y0}\,\cos\phikptQ\right).\\
\end{align}
The structure of the splitting function and the meaning of the different pieces is the same as in the PS meson case in Eq. (\ref{eq:splitting F_PS explicit}). The differences are the introduction of the invariant mass distribution $|D_{\h}(\mh^2)|^2$ and the amplitudes of the modulations in the azimuthal angle $\phikptQ$ in the last line. As shown in Ref. \cite{Kerbizi:2021M20}, for VM production the amplitudes of the modulations in the distribution of $\phikptQ$ are the opposite to those of the PS case, and they are reduced by the factor $\fL$ (cf. with the last line in Eq. (\ref{eq:splitting F_PS explicit})).

\subsubsection{The polarized decay of the vector meson}\label{sec:decays}
\begin{figure}[tb]
\centering
\begin{minipage}[b]{0.5\textwidth}
\includegraphics[width=0.65\textwidth]{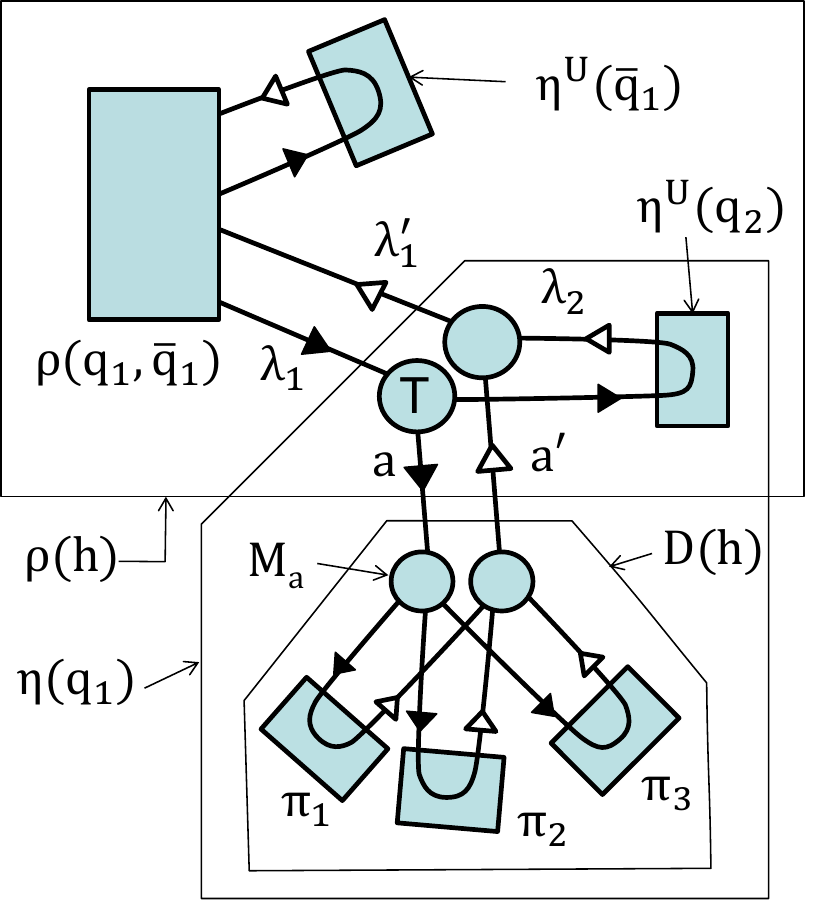}
\end{minipage}
\caption{Unitarity diagram for $\q\qbar\rightarrow hX$, where $h$ is a VM emitted from $\q$ and decays in three pions. Particularly is shown the decay matrix of the VM.}
\label{fig:rho(h)}
\end{figure}

The VM has been emitted. Now we consider its decay. The spin density matrix of $h$ can be calculated by inserting Eq. (\ref{eq:eta(q) VM}) in Eq. (\ref{eq:A^2 e+e- simplified}), imposing $\eta(\qbar)=\Iqbar$ and $\eta(\qp)=\Iq$, and freeing the polarization indices of the meson. The result is
\begin{align}\label{eq:rho(h) quark}
\nonumber    \rho_{aa'}(h) &= \frac{\rm\, \Tr_{\qp\qbar}\left[\T^{a}_{\qp,h=\VM,\q}\,\rho(\q,\qbar)\,\T^{\dagger\,a'}_{\qp,h=\VM,\q} \right]}{\Tr\left[\dots\right]}
\\ 
&= \frac{\Cqq_{\alpha 0}\rm\, \Trq\left[\Delta(\kpt)\,\Gammaq^a(h)\,\sigmaq_{\alpha}\,\Gammaq^{a'\,\dagger}(h)\,\Delta^{\dagger}(\kpt)\right]}{\Tr\left[\dots\right]}
\end{align}
and is represented by the upper rectangular domain in Fig. \ref{fig:rho(h)}. In the denominator $\left[\dots\right]$ indicates the same expression as in the numerator but the trace is taken also on the polarization index $a$ of the VM. $a$ and $a'$ can take the values $x,y,z$ and span the polarization states of the meson measured in the meson rest frame~\footnote{This rest frame is obtained from the c.m.s by the boost composition of Ref. \cite{Kerbizi:2021M20}.}. We have also used the vector of couplings matrices $\Gammaq^a(h)=(\GT\sigmaXq\sigmaZq,\GT\sigmaYq\sigmaZq,\GL\Iq)$ (see Eq.(\ref{eq:Gamma q})).

The spin density matrix of the VM is used for the generation of the anisotropic angular distribution of the hadrons produced in the decay of the meson. Indicating with $\mathcal{M}_a(p_1,p_2,\dots)$ the matrix element that describes the decay process $\VM\rightarrow p_1,p_2,\dots$, the angular distribution of the decay products in the rest frame of the VM can be obtained by
\begin{align}\label{eq:decay distribution}
    \frac{d\mathcal{N}}{d\Phi(p_1,p_2,\dots)} = \frac{\mathcal{M}_a(p_1,p_2,\dots)\,\rho_{aa'}(h)\,\mathcal{M}_{a'}^{\dagger}(p_1,p_2,\dots)}{\int\,d\Phi(p_1,p_2\,\dots)\,[\dots]}.
\end{align}
The quantity $d\Phi(p_1,p_2,\dots)$ indicates the differential phase space element involved in the decay. The detailed description of the decay processes included in the string+${}^3P_0$ model is given in Ref. \cite{Kerbizi:2021M20}. The quantities in the numerator on the RHS of Eq. (\ref{eq:decay distribution}) are represented in Fig. \ref{fig:rho(h)} for the example $h=\omega\rightarrow 3\pi$.

Following the CK recipe, the decay process of the VM fixes the decay matrix $D(\hat{p}_1,\hat{p}_2,\dots)$, where $\hat{p}_1,\hat{p}_2,\dots$ are the generated four-momenta of the decay hadrons. The decay matrix is necessary to account for the quantum mechanical correlations between the orientation of the decay hadrons and the spin state of the leftover quark $\qp$ or the antiquark $\qbar$. It is given by
\begin{align}\label{eq:decay matrix}
    D_{a'a}(\hat{p}_1,\hat{p}_2,\dots) = \mathcal{M}^{\dagger}_{a'}(\hat{p}_1,\hat{p}_2,\dots)\,\mathcal{M}_a(\hat{p}_1,\hat{p}_2,\dots),
\end{align}
hence it is evaluated at the generated momenta of the decay hadrons and depends on the matrix element that describes the decay process. The formation of the decay matrix is represented in Fig. \ref{fig:rho(h)} by the hexagonal domain. 

\subsubsection{Propagation of the spin correlations}\label{sec:propagation of correlations}
After the emission of the hadron $h$ from the quark end of the string by the splitting $\q\rightarrow h + \qp$ the string piece stretched between $\qp$ and $\qbar$ remains to be fragmented. 
The $\qp\qbar$ string is characterized by a new joint spin density matrix $\rho(\qp,\qbar)$ and associated new correlation coefficients $\C^{\qp\qbar}$. 
If $h=\PS$ the joint spin density matrix can be calculated by inserting in Eq. (\ref{eq:A^2 e+e- simplified}) the acceptance matrix $\eta(\q)$ in Eq. (\ref{eq:acceptace q and qbar matrix form}) and identifying the resulting expression with $d\sigma\propto \Tr_{\qp\qbar}\left[\rho(\qp,\qbar)\,\eta(\qp)\otimes\eta(\qbar)\right]$. If $h=\VM$, the acceptance matrix to be used is that in Eq. (\ref{eq:eta(q) VM}) with the decay matrix in Eq. (\ref{eq:decay matrix}).
We obtain (given the emission of the hadron $h$ from the quark end of the string) for the joint spin density matrix of the system $\qp\qbar$
\begin{equation}\label{eq:density matrix q'qbar}
\begin{aligned}
    \rho(\qp,\qbar) &= \begin{cases}
        \T_{\qp,h,\q} \,\rho(\q,\qbar)\,\T^{\dagger}_{\qp,h,\q}/\Tr\left[\dots \right] & \rm{h=\PS}\\
        \T^a_{\qp,h,\q} \,\rho(\q,\qbar)\,\T^{\dagger\,a'}_{\qp,h,\q}\,D_{aa'}/\Tr\left[\dots\right] & \rm{h=\VM}
    \end{cases}.
\end{aligned}
\end{equation}
It is represented by the domain C in Fig. \ref{fig:amplitude squared e+e-}. The associated correlation coefficients are
\begin{align}\label{eq:Cq'qbar}
\nonumber    \C^{\qp\qbar}_{\alpha'\beta} &= \Tr_{\qp\qbar}\left(\rho(\qp,\qbar)\,\sigma_{\alpha'}(q)\otimes\sigma_{\beta}(\qbar)\right)\\
    &= \Cqq_{\alpha\beta}\,M^{\q}_{\alpha\alpha'}\,/\,\Cqq_{\alpha 0}\,M^{\q}_{\alpha 0}.
\end{align}
Thus the correlation coefficients $\C^{\qp\qbar}$ can be obtained by matrix operations on the correlation coefficients $C^{\q\qbar}$ by introducing the matrix $M^{\q}_{\alpha\alpha'}$. Such matrix for PS and VM emissions is given by
\begin{equation}\label{eq:M matrices quark}
    \begin{aligned}
        &M^{\q}_{\alpha\alpha'}|_{\PS}  = \frac{1}{2}
        \Tr\left[ \sigmaq_{\alpha'}\,\Delta(\kpt)\,\Gammaq_{h}\,\sigmaq_{\alpha}\,\GammaqDag_h\,\Delta^{\dagger}(\kpt)\right], \\
        &M^{\q}_{\alpha\alpha'}|_{\VM} = \frac{1}{2}\Tr\left[ \sigmaq_{\alpha'}\,\Delta(\kpt)\,\Gammaq^a(h)\,\sigmaq_{\alpha}\,\Gammaq^{\dagger\,a'}(h)\,\Delta^{\dagger}(\kpt)\right]\,D_{a'a}.
    \end{aligned}
\end{equation}

\subsubsection{Hadron emission from the antiquark end}
\paragraph{Pseudoscalar meson emission.}
The emission of a hadron $H$ from the antiquark end of the string by the splitting $\qbar\rightarrow H+\qpbar$, after the hadron $h$ was emitted from the quark end proceeds in a way symmetrical to that of the emission of $h$. The emission of $h$, which has already occurred, has changed the $\q-\qbar$ string in the $\qp-\qbar$ one and we must take into account the spin information coming from the splitting $\q\rightarrow h+\qp$. This information is contained in the spin density matrix $\rho(\qp,\qbar)$ of Eq. (\ref{eq:density matrix q'qbar}). Therefore, to get the momentum spectrum of $H$ in the PS case it suffices to make in Eq. (\ref{eq:splitting F_PS}) the replacement $\q\rightarrow \qbar$, $\qbar\rightarrow \qp$ and $\qp\rightarrow \qpbar$, $h\rightarrow H$, $Z_+\rightarrow Z_-$ and $\pt\rightarrow \Pt$. 
We obtain
\begin{align}\label{eq:splitting F_PS qbar}
\nonumber  &\frac{dP(\qbar\rightarrow H=\PS + \qpbar; \qp\qbar)}{dZ_-Z_-^{-1}\,d^2\Pt} \equiv F_{\qpbar,H,\qbar}(Z_-,\Pt;\ktbar,\C^{\qp\qbar}) \\
&= \rm{Tr}_{\qp\qpbar}\left[\T_{\qpbar,H,\qbar}\,\rho(\qp,\qbar)\, \T_{\qpbar,H,\qbar}^{\dagger} \right],
\end{align}
where we have defined $\T_{\qpbar,H,\qbar}=\Iqp\otimes T_{\qpbar,H,\qbar}$, and the joint spin density matrix $\rho(\qp,\qbar)$ is given in Eq. (\ref{eq:density matrix q'qbar}). The splitting function $F_{\qpbar,H=\PS,\qbar}$ depends on $Z_-$, $\Pt$, $\ktbar$ and on the correlation coefficients $C^{\qp\qbar}$, calculated in Eq. (\ref{eq:Cq'qbar}).
It gives the conditional probability of emitting the hadron $H$, given that the hadron $h$ was emitted from the quark end.


Inserting in Eq. (\ref{eq:splitting F_PS qbar}) the splitting matrix in Eq. (\ref{eq:T antiquark}), the explicit expression for $F_{\qpbar,H=\PS,\qbar}$ is
\begin{align}\label{eq:splitting F_PS explicit qbar}
\nonumber    F&_{\qpbar,H=\PS,\qbar}(Z_-,\Pt;\ktbar,\C^{\qp\qbar})
= \frac{\hat{u}_{\qbar,H}}{\hat{u}_{\qbar}}\, \left(\frac{1-Z_-}{\ehbar^2}\right)^{a}\,e^{-\bL\ehbar^2/Z_-}\,\\
\nonumber &\times N_a^{-1}(\ehbar^2)\, \fT^2(\kptbarkptbar)\,\frac{|\mu|^2+\kptbarkptbar}{|\mu|^2+\langle \kTkT\rangle_{\fT}}\\
&\times \left(1+\hat{a}(\kptabsbar)\,\C^{\qp\qbar}_{0x}\,\sin\phikptQbar-\hat{\mathnormal{a}}(\kptabsbar)\,\C^{\qp\qbar}_{0y}\,\cos\phikptQbar\right).
\end{align}
This expression is similar to Eq. (\ref{eq:splitting F_PS explicit}), with the substitutions $Z_+\rightarrow Z_-$, $\lbrace \kt, \pt, \kpt\rbrace\rightarrow\lbrace \ktbar, \Pt, \kptbar \rbrace$, $\phikptQ\rightarrow \phikptQbar$ and $\C^{\q\qbar}\rightarrow \C^{\qp\qbar}$. The azimuthal angle $\phikptQbar$ of the transverse momentum $\kptbar$ of $\qpbar$ is measured in the AHF. The azimuthal angle in the QHF can be obtained using Eq. (\ref{eq:angles}). 
Equation (\ref{eq:splitting F_PS explicit qbar}) is responsible for a Collins effect for the production of $H$ in the AHF. The strength of the effect depends on the emission of $h$ from the quark side through the correlation coefficients $\C^{\qp\qbar}$. A similar Collins effect would have been found if we had treated the emission of $h$ after that of $H$. In fact, these effects are related and they sum up to an azimuthal correlation between $\pt$ and $\Pt$ given by Eq. (\ref{eq:P(e+e- -> h H X) final}).

\paragraph{Vector meson emission.} The splitting function for the emission of a VM in the antiquark splitting $\qbar\rightarrow H=\VM + \qpbar$ can be obtained from Eq. (\ref{eq:splitting F_PS qbar}) by the substitution $\Iq\otimes T_{\qpbar,h=\PS,\qbar}\otimes \rightarrow \Iq\otimes T^a_{\qpbar,h=\VM,\qbar}\,\V_a$ and by summing over the polarization states of the meson. The corresponding splitting function takes the explicit form
\begin{equation}\label{eq:splitting F_VM explicit antiquark}
\begin{aligned}
&F_{\qpbar,H=\VM,\qbar}(\mh^2,Z_-,\Pt;\ktbar,\C^{\qp\qbar})
= \frac{\hat{u}_{\qbar,H}}{\hat{u}_{\qbar}}\, |D_{\hbar}(\mh^2)|^2\,\\
 &\times \left(\frac{1-Z_-}{\ehbar^2}\right)^{a}\,e^{-\bL\ehbar^2/Z_-}\, N_a^{-1}(\ehbar^2)\, \\
 &\times \fT^2(\kptbarkptbar)\,\frac{|\mu|^2+\kptbarkptbar}{|\mu|^2+\langle \kTkT\rangle_{\fT}}\\
 &\times \left(1-\hat{a}(\kptabsbar)\,\fL\,\C^{\qp\qbar}_{0x}\,\sin\phikptQbar+\hat{\mathnormal{a}}(\kptabsbar)\,\fL\,\C^{\qp\qbar}_{0y}\,\cos\phikptQbar\right).
\end{aligned}
\end{equation}
We note again that the azimuthal angle $\phikptQbar$ of the transverse momentum of $\qpbar$ is measured in the AHF.

The density matrix of $H$ as well as the correlation coefficients of the new string piece after the antiquark splitting $\qbar\rightarrow H + \qpbar$ have similar expressions to those that are obtained when the hadron is emitted from the quark side (see Sec. \ref{sec:decays} and Sec. \ref{sec:propagation of correlations}). The calculation is not repeated here and the explicit expressions can be found in Appendix \ref{app:VM emission antiquark}.

\begin{figure}[tbh]
\centering
\begin{minipage}[b]{0.50\textwidth}
\includegraphics[width=0.99\textwidth]{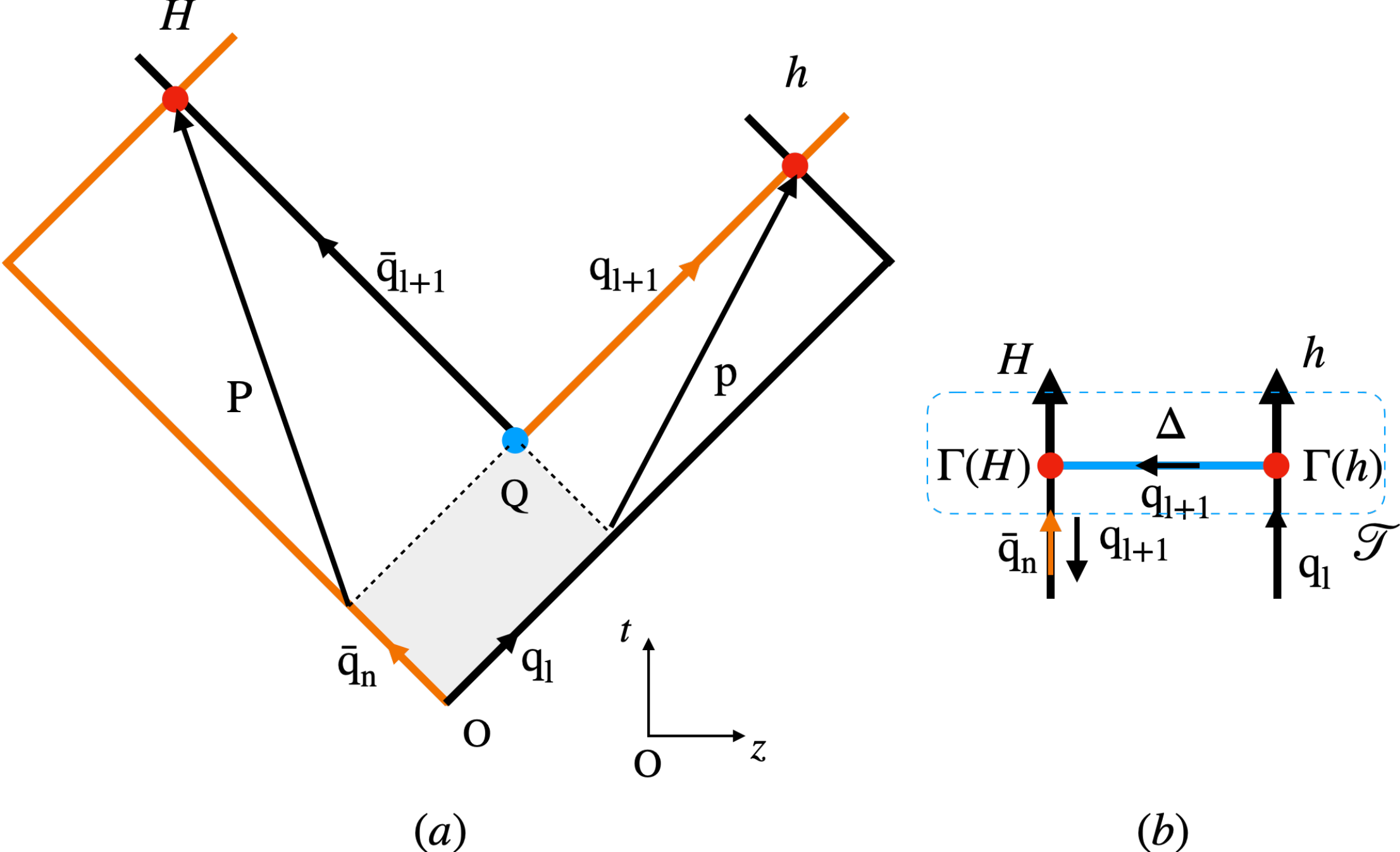}
\end{minipage}
\caption{(a) Spacetime picture of the string fragmentation of the $\ql\qlbar$ pair and the production of the last two hadrons $h$ and $H$. (b) Associated amplitude for the two-step process $\ql\rightarrow h + q_{l+1}$ and $q_{l+1}\rightarrow H+q_{l+2}$.}
\label{fig:exit condition}
\end{figure}

\subsubsection{End of the fragmentation chain}\label{sec:end of the chain}
We assume that several hadrons have been emitted from both string ends, and that a string piece stretched between a quark $\ql$ and an antiquark $\qlbar$ remains to be fragmented by the process $\ql\qlbar\rightarrow h + H$. Following the recipe adopted in the MC implementation of the Lund Model of string fragmentation \cite{Andersson:1983ia}, the condition for the termination of the fragmentation chain is called if the squared mass of the $\ql\qlbar$ string falls below some minimum value $M_{\rm min}^2$ of the order of a $\GeV$ (for the precise definition of $M_{\rm min}^2$ see Ref. \cite{Andersson:1983ia}). The fragmentation chain is thus ended by the creation of a last quark pair and the formation of the hadrons $h$ and $H$. The corresponding spacetime picture is shown in Fig.\ref{fig:exit condition}a.

To construct these two hadrons we use the following recipe. We consider the reaction  $q_{l}+\bar q_n \to  h+H$ equally as $q_{l}\to h + q_{l+1} $ followed by $q_{l+1}\to H + q_{l+2}$, as shown in Fig. \ref{fig:exit condition}b, or as $\bar q_n \to  \bar H + q_{n+1}$ followed by  $\bar q_{n+1}\to h + q_{n+2}$, where $\bar q_n$, $\bar q_{n+1}$ and $\bar q_{n+2}$ are respectively the antiparticles of $q_{l+2}, q_{l+1}$ and $q_{l}$. The $\ql\qlbar$ pair is characterized by the joint spin density matrix $\rho (q_{l},\bar q_n) = 4^{-1} \, C^{\ql\qlbar}_{\mu\nu} \,  \sigma^{q_l}_{\mu} \otimes \sigma^{\bar q_n}_{\nu}$. If $\ql$ was generated after $\qlbar$, Eqs. (\ref{eq:density matrix q'qbar}-\ref{eq:Cq'qbar}) are used to calculate such matrix; otherwise, Eqs. (\ref{eq:density matrix q'qbar'}-\ref{eq:Cq'qbar'}) are used. $q_{l}$, $\mu$, $\bar q_n$, $\nu$ take the place of $q_2$, $\alpha'$,  $\bar q_1$, $\beta$ in the first case, of $q_2$, $\alpha$,  $\bar q_2$ and $\beta'$ in the second case. 

Starting with $\rho (q_{l},\bar q_n)$, we build the matrix $R(q_{l},q_{l+2})$ obtained by the successive transformations:
\begin{itemize}
\setlength\itemsep{0em}
\item[a)] change the signs of $C^{\ql\qlbar}_{\mu y}$ and $C^{\ql\qlbar}_{\mu z}$ (to put everything in the QHF frame),
\item[b)] reverse the signs of $C^{\ql\qlbar}_{\mu i}$ for $i=x,y$ or $z$, 
\item[c)] multiply the resulting matrix on the left and on the right by $\Iden \otimes \sigma_z$\footnote{ 
 Steps b) and c) come from the expression $\xi(\Sv) = \sigma_z\, \chi(-\Sv)$ of an antiquark spinor in the string+$^3P_0$ model \cite{Kerbizi:2018qpp}, where $\chi(\Sv)$ and $\xi(\Sv)$ are the 2-D reductions of the Dirac spinors $u(\pv,\Sv)$ and $v(\pv,\Sv) = \gamma_5  u(\pv,-\Sv)$ in the $\langle\alpha_z\rangle=+1$ subspace. The choice of this subspace is motivated by the relation $\langle \alpha_z\rangle =v_z$ and the fact that in Fig. \ref{fig:exit condition} a longitudinal velocity $v_z=+1$ is attributed to the antiquark just before it reaches the emission vertex of $H$.}.
 \end{itemize}
 
To fix the hadron $h$ and $H$, we draw the flavor $u$, $d$ or $s$ of $q_{l+1}$  with respective probabilities proportional to
$ \hat u_{\rm u}$, $\hat u_{\rm d}$ or $\hat u_{\rm s}$ (see Eq. (\ref{eq:u matrix})). Then the hadron species are drawn with probabilities proportional to $|C_{q_{l+1},h, q_{l}}|^2$ and $|C_{q_{l+2},H, q_{l+1}}|^2$. The mass of VMs is generated according to the corresponding distribution $|D(M^2)|^2$. 

To build the four-momenta of $p$ of $h$ and $P$ of $H$, it is necessary to know the transverse momentum $\kT(q_{l+1}) $ of $q_{l+1} $. The latter is generated with the probability proportional to
\be
\langle j  | {\cal T}^{a,b}  | i  \rangle \, \langle i \otimes j' | R |i' \otimes j \rangle \, \langle j'  | {\cal T}^{\dag a,b} | i'  \rangle
\ee
(summed over repeated indices), where $i$, $j$, $a$ and $b$ are spin states of $q_{l}$, $q_{l+2}$, $h$ and $H$, and  
\be
 {\cal T}^{a,b} =\Gamma^b(H) \, \Delta(\kT(q_{l+1})) \, \Gamma^a(h). 
\ee
The latter amplitude is also shown in Fig. \ref{fig:exit condition}b. The transverse momenta of the emitted hadrons then calculated as $\pt=\kT(q_{l}) - \kT(q_{l+1}) $ and  $\Pt=\kT(q_{l+1}) - \kT(q_{l+2}) $.

Finally, the longitudinal momenta (and the energies) of the final two hadrons are obtained by calculating the two possible solutions of the system
\begin{eqnarray}
\nonumber && p^+ \, p^- = m_h^2+ \ptpt, \\
&& (P_{\rm rem}^+-  p^+) \, (P_{\rm rem}^--p^-) = m_H^2+ \PtPt, 
\end{eqnarray}
with $P_{\rm rem}=k(q_{l}) + k(\qlbar)=k(\ql)-k(q_{l+2})$ and choosing one of them, each having a relative  weight  proportional to $\exp\{-b_L \, P^+ \, p^-\}$. The factor $P^+p^-$ is the area in the past lightcone of the spacetime point $Q$ in Fig. \ref{fig:exit condition}a. The exponential gives the probability that no string breaking occurred in the past lightcone of $Q$. 

If $h$ and $H$ are VMs, their spin states are described by the joint density matrix of $h$ and $H$
\be
\langle a,b |\rho(h,H) | a', b'\rangle \propto 
\langle j  | {\cal T}^{a,b}  | i  \rangle  \langle i \otimes j' | R |i' \otimes j  \rangle  \langle j'  | {\cal T}^{\dag a',b'} | i'  \rangle.
\ee
If only $h$ is a VM, one gets the single density matrix $\langle a |\rho(h) | a'\rangle$ by omitting the indices $b$ and $b'$. The case where only $H$ is a VM is analogous. 
If $h$ is a VM, its decay is treated like in Sec. \ref{sec:decays} with $\rho(h)=\Tr_H\,  \rho(h,H)$. If $H$ is also a VM, to generate its decay one must first calculate the decay matrix $D(h)$ as in Sec. \ref{sec:decays} according to the CK recipe, and then decay $H$ according to the density matrix
\be 
\langle b |\rho(H) | b'\rangle \propto \langle a,b |\rho(h,H) | a', b'\rangle \, \langle a' |D(h) | a\rangle. 
\ee

The proposed recipe is formulated at the amplitude level, and it can be regarded as the generalization of the recipe of the Lund Model \cite{Andersson:1983ia} employed for the joining of the quark and antiquark jets in Monte Carlo simulations.

\subsection{The recursive recipe for the string fragmentation}\label{sec:recursive recipe}
By gathering the ingredients presented in Sec. \ref{sec:steps}, we can now formulate the following recipe for the fragmentation of a string stretched between a quark $\q$ and an antiquark $\qbar$ with entangled spin states. As already mentioned, we assume the $\q\qbar$ pair to be produced in the hard process $e^+e^-\rightarrow \gamma^*\rightarrow \q\qbar$. The recipe is however general and can be applied to other processes as well. In fact, the information on the hard process that produces the pair is included only in the joint spin-density matrix $\rho(\q,\qbar)$, which is required as an initial condition for the simulation of the hadronization process $\q\qbar\rightarrow h_1,h_2,\dots, h_N$.

We divide the simulation of the $e^+e^-$ annihilation in the generation of the primary quark pair and in the string fragmentation of the pair.

\textbf{Generation of the primary quark pair.}
Generate the flavor of $\q$ using the relative probabilities $\hat{P}(u\bar{u}):\hat{P}(d\bar{d}):\hat{P}(s\bar{s}):\hat{P}(c\bar{c}):\dots$, where $\hat{P}(\q\qbar)=\hat{\sigma}(\q\qbar)/\sum_{q}\hat{\sigma}(q\bar{q})$ and $\hat{\sigma}=\int\,d\cos\theta\,d\hat{\sigma}/d\cos\theta$ is the integrated hard cross section obtained using Eq. (\ref{eq:hard cross section}). Then generate the polar angle $\theta$ using the probability distribution obtained by the cross section ratio $\hat{\sigma}^{-1}\,d\hat{\sigma}/d\hat{\Omega}$. Set up the four-momenta of $e^+$, $e^-$, $\q$ and $\qbar$ using Eq. (\ref{eq:momenta}) and use the generated value of $\theta$ to calculate the correlation coefficients $\Cqq$. The latter can be read from the expression of the joint spin density matrix of the $\q\qbar$ pair in Eq. (\ref{eq:joint density matrix}).

\textbf{String fragmentation of the quark pair.} The initial conditions for the string fragmentation process are the momenta $\k$ of $\q$ and $\kbar$ of $\qbar$ and the joint density matrix $\rho(\q,\qbar)$. From the quark momenta we define the available lightcone momenta $\Ptot^+=\sqrt{s}=\Ptot^-$, where $\Ptot=\k+\kbar$. It is $\kt=\ktbar=\textbf{0}$, and thus $\PtotT=\textbf{0}$. To generate recursively the string fragmentation process of the $\q\qbar$ pair, repeat the following steps:
\begin{itemize}
    \item [1.] Select with equal probability whether to emit the first hadron $h$ from the quark side or the antiquark side\footnote{As done in the Lund string Model \cite{Andersson:1983ia}.}. 
    \item[2.] \textbf{If the splitting is performed from the quark side:} select a new quark pair $\qp\qpbar$ with probability $\hat{u}_{\q,h}/\hat{u}_{\q}$ using Eq.~(\ref{eq:u matrix}), form the hadron $h=(\q\qpbar)$ and decide whether it is a VM with probability $\fVM$, or a PS meson. If $h$ is a PS meson, generate $\kptkpt$, $\phikptQ$ and $Z_+$ using the splitting function $F_{\qp,h=\PS,\q}(Z_+,\pt;\kt,\Cqq)$ in Eq. (\ref{eq:splitting F_PS explicit}). If $h$ is a VM use instead the splitting function $F_{\qp,h=\VM,\q}(\mh^2,Z_+,\pt;\kt,\Cqq)$ in Eq. (\ref{eq:splitting F_VM explicit}) to generate first the invariant mass squared $\mh^2$, and then $\kptkpt$, $\phikptQ$ and $Z_+$. 
    Calculate the lightcone momenta $p^{+}=Z_+\k^+$ and $p^{-}=\eh^2/p^+$, the new available light-cone momenta $(\Ptot^+)^{\rm new}=\Ptot-p^+$ and $(\Ptot^-)^{\rm new}=\Ptot^--\eh^2/p^+$, and the transverse momentum $\PtotT^{\rm new}=\PtotT-\pt$. If $(\Ptot^2)^{\rm new}<M_{\rm min}^2$, go to step 4. Otherwise set $\Ptot = (\Ptot)^{\rm new}$ and continue by constructing the four-momentum of $h$ using $p=(E_h,\pt,p_L)$, where $E_h=(p^++p^-)/2$ and $p_L=(p^+-p^-)/2$.

    If $h$ is a VM apply the following further steps to decay the meson:
        \begin{itemize}
            \item [2.1] Calculate the spin density matrix of $h$ using Eq. (\ref{eq:rho(h) quark}), and generate the momenta of the decay hadrons in the rest frame of $h$ using Eq. (\ref{eq:decay distribution}). The expressions for the decay amplitude $\mathcal{M}_a$ can be found in Ref. \cite{Kerbizi:2021M20}.
            \item [2.2] Calculate the decay matrix $D$ using Eq. (\ref{eq:decay matrix}).
            \item [2.3] To come back to the center of mass frame, apply the composition of longitudinal and transverse boosts in Ref. \cite{Kerbizi:2021M20} to the decay hadrons.
        \end{itemize}
    
    
    \item[3.] Calculate the correlation coefficients $\C^{\qp\qbar}$ of the new string piece with endpoints $\qp$ and $\qbar$ using Eqs. (\ref{eq:Cq'qbar})-(\ref{eq:M matrices quark}). 
    Let $\qp$, $\kp$ and $\C^{\qp\qbar}$ take the place of $\q$, $\k$ and $\C^{\q\qbar}$, respectively, and go to step 1.

    \textbf{If the splitting is performed from the antiquark side:} the steps are similar to the splitting from the quark side, and can be found in Appendix \ref{sec:steps antiquark splitting}.
    
    \item[4.] The mass of the remaining string piece $\ql-\qlbar$ has become less than $M_{\rm min}$. To terminate the fragmentation chain hadronize the remaining $\ql\qlbar$ pair by generating the last quark pair $q_{l+1}\bar{q}_{l+1}$, and forming the hadrons $h=(\ql\bar{q}_{l+1})$ and $H=(\qlbar q_{l+1})$. For each hadron, decide whether it is a VM with probability $\fVM$, or a PS meson. Generate the longitudinal splitting variable and transverse momentum of $q$ by the recipe in Sec. \ref{sec:end of the chain}. Calculate the transverse momenta $\pt=\klt-\kT(q_{l+1})$ and $\Pt=\kltbar+\kT(q_{l+1})$. 
    Finally build the four-momentum $p$ of $h$ and $P$ of $H$.
\end{itemize}

The steps 1-4 are similar to those applied for the implementation of the Lund string Model in the PYTHIA event generator \cite{Sjostrand:2007gs,Bierlich:2022pfr}. In addition the CK recipe and the rules of the string+${}^3P_0$ model are used 
to account for the spin correlations at each hadron emission, and to propagate these correlations after each emission as required by quantum mechanics. This recursive recipe is therefore suitable for the implementation of the $e^+e^-$ annihilation with spin effects in MCEGs. The natural choice would be the implementation in PYTHIA 8 \cite{Bierlich:2022pfr} by extending the StringSpinner package \cite{Kerbizi:2023cde}, which currently is applied only to the polarized SIDIS process.




\section{Application to back-to-back hadron production in $e^+e^-$}\label{sec:application}
The recipe for the fragmentation of a string stretched between a quark pair with correlated spin states described in Sec. \ref{sec:steps} can be checked to reproduce the expected azimuthal distribution by applying it to the process $e^+e^-\rightarrow h\,H\,X$. As in Sec. \ref{sec:steps}, we assume $h$ to be emitted in the splitting $\q\rightarrow h+\qp$, and $H$ to be emitted in the splitting $\qbar\rightarrow H +\qpbar$. The two hadrons are associated to different quark jets and are thus expected to be produced nearly back-to-back in the c.m.s. The calculations are shown in Sec. \ref{sec:back-to-back PS PS} for the case $h=\PS$ and $H=\PS$, and in Sec. \ref{sec:back-to-back VM PS} for the case $h=\VM$ and $H=\PS$. For these calculations we neglect the quark mass $m_q$.

\subsection{Production of back-to-back $\PS$ mesons}\label{sec:back-to-back PS PS}
According to the recipe described in Sec. \ref{sec:steps}, the probability of producing the two hadrons $h$ and $H$ is obtained as a three step process (e.g., starting the string fragmentation from the $\q$ side): i) the production of the $\q\qbar$ pair in the hard process $e^+e^-\rightarrow \q\qbar$, ii) the splitting $\q\rightarrow h +\qp$ given the $\Cqq$ correlation coefficients and iii) the splitting $\qbar\rightarrow H+\qpbar$ given the correlation coefficients $\C^{\qp\qbar}$. The probability for the step i) to occur is given by the cross section for the hard scattering. The probabilities for the steps ii) and iii) to occur are given by the splitting functions of Eqs. (\ref{eq:splitting F_PS explicit}) and (\ref{eq:splitting F_PS explicit qbar}), respectively. The total probability is the product of the three probabilities, and it can be written as (cf. Eq. (\ref{eq:A^2 e+e-}) with $\eta(\qp)=\Iqp$ and $\eta(\qpbar)=\Iden^{\qpbar}$) 
\begin{eqnarray}\label{eq:e+e- -> h H X calculated}
    \begin{aligned}
        & dP(e^+e^-\rightarrow h\,H\,X)=\hat{\sigma}^{-1}\frac{d\hat{\sigma}}{d\cos\theta} \,d\cos\theta \\
        &\times F_{\qp,h,\q}(Z_+,\pt;\kt,\Cqq)\,Z_+^{-1}dZ_+\,d^2\pt \\
        &\times F_{\qpbar,H,\qbar}(Z_-,\Pt;\ktbar,\C^{\qp\qbar})\,Z_-^{-1}dZ_-d^2\Pt,
    \end{aligned}
\end{eqnarray}
where $\kt = 0$ and $\ktbar = 0$.

To calculate the splitting function $F_{\qp,h,\q}(Z_+,\pt;\kt,\Cqq)$ of Eq. (\ref{eq:splitting F_PS explicit}) one takes $\Cqq_{x0}=\Cqq_{y0}=0$, according to Eq. (\ref{eq:initial C coefficients}). Therefore $h$ is emitted with a flat azimuthal distribution. For the splitting function $F_{\qpbar,H,\qbar}$ of Eq. (\ref{eq:splitting F_PS explicit qbar}), instead, the coefficients $\C^{\qp\qbar}_{0x}$ and $\C^{\qp\qbar}_{0y}$ are needed. They can be calculated by using Eqs. (\ref{eq:Cq'qbar})-(\ref{eq:M matrices quark}), with the explicit expressions for the splitting amplitude in Eq. (\ref{eq:T quark}) and the quark coupling to PS mesons in Eq. (\ref{eq:Gamma q}). We obtain 
\begin{align}\label{eq:Cq'qbar calculated}
\nonumber    \C^{\qp\qbar}_{0x} &= \C^{\q\qbar}_{xx}\,\frac{M^{\q}_{x0}}{M^{\q}_{00}}= -\frac{\sin^2\theta}{1+\cos^2\theta}\,\hat{a}(\kptabs)\,(-\sin\phikptQ), \\
    \C^{\qp\qbar}_{0y} &= \C^{\q\qbar}_{yy}\,\frac{M^{\q}_{y0}}{M^{\q}_{00}}= -\frac{\sin^2\theta}{1+\cos^2\theta}\,\hat{a}(\kptabs)\,\cos\phikptQ,
\end{align}
meaning that the quark $\qbar$ has a transverse polarization that depends on the transverse momentum of $h$. According to Eq. (\ref{eq:splitting F_PS explicit qbar}), this means that $H$ is emitted with a Collins effect in the AHF.

The hadron transverse momenta are given by $\pt=-\kpt$ and $\Pt=-\kptbar$. Indicating by $\phi^{q}_h$ and $\phi^{q}_H$ the azimuthal angles of $\pt$ and $\Pt$ in the helicity frame of $q=\q,\qbar$, respectively, it is $\phi^{\q}_h=\phikptQ+\pi$ and $\phi^{\qbar}_H=\phikptQbar+\pi$. The azimuthal angle $\phi^{\qbar}_H$ can be expressed in the QHF using Eq. (\ref{eq:angles}), and it is $\phi_H^{\q}=\pi-\phikptQbar$.

With these considerations, and inserting in Eq. (\ref{eq:e+e- -> h H X calculated}) the Eq. (\ref{eq:splitting F_PS explicit}) and the Eq. (\ref{eq:splitting F_PS explicit qbar}) with the coefficients in Eq. (\ref{eq:Cq'qbar calculated}), we obtain the probability for the PS + PS case
\begin{widetext}
\begin{eqnarray}\label{eq:P(e+e- -> h H X) final}
    \begin{aligned}
        &\frac{dP(e^+e^-\rightarrow h\,H\,X)}{d\cos\theta\,dZ_+\,d^2\pt\,dZ_-\,d^2\Pt} = \frac{3}{8}\,(1+\cos^2\theta) \times \frac{\hat{u}_{\q,h}}{\hat{u}_{\q}}\, Z_+^{-1}\,\left(\frac{1-Z_+}{\eh^2}\right)^{a}\,e^{-\bL\eh^2/Z_+}\,N_a^{-1}(\eh^2)\, \fT^2(\ptpt)\,\frac{|\mu|^2+\ptpt}{|\mu|^2+\langle \ptpt\rangle_{\fT}}\\
        &\times \frac{\hat{u}_{\qbar,H}}{\hat{u}_{\qbar}}\,Z_-^{-1}\, \left(\frac{1-Z_-}{\ehbar^2}\right)^{a}\,e^{-\bL\ehbar^2/Z_-}\,N_a^{-1}(\ehbar^2)\, \fT^2(\PtPt)\,\frac{|\mu|^2+\PtPt}{|\mu|^2+\langle \PtPt\rangle_{\fT}}
        \times \left(1+\frac{\sin^2\theta}{1+\cos^2\theta}\,\hat{a}(p_{\rm T})\,\hat{a}(P_{\rm T})\,\cos(\phi^{\q}_h+\phi^{\q}_H)\right).
    \end{aligned}
\end{eqnarray}
\end{widetext}
As can be seen from the last factor of the equation, the recipe produces the azimuthal modulation $\cos(\phi_h^{\q}+\phi_H^{\q})$ associated to the Collins asymmetry for the production of two back-to-back hadrons in $e^+e^-$, as calculated in Refs. \cite{Boer:2008fr,DAlesio:2021dcx}. Also, the amplitude of the modulation has a positive sign, as observed by the BELLE \cite{Belle:2011cur} and BABAR \cite{BaBar:2013jdt,BaBar:2015mcn} experiments. The azimuthal angles of both hadrons are referred to the same reference system, which in this case is the QHF. The amplitude of the modulation is proportional to the squared imaginary part $(\rm{Im}\,\mu)^2$ of the complex mass $\mu$. If the $\rm{Im}\mu$ vanishes then the asymmetry vanishes, as is the case for the transverse spin effects in the string+${}^3P_0$ model \cite{Kerbizi:2018qpp,Kerbizi:2021M20}.

The simple formula (\ref{eq:P(e+e- -> h H X) final}) holds only for the two leading hadrons produced in each quark jet. In order to obtain the complete results for the Collins asymmetries in $e^+e^-$ annihilation, the present model must be implemented in a Monte Carlo event generator, either standalone as in Ref. \cite{Kerbizi:2021M20} or in the PYTHIA MCEG by extending the StringSpinner package in Ref. \cite{Kerbizi:2023cde}. Still, this calculation is important as it demonstrates that the recipe presented in this work reproduces the azimuthal correlation of the hadrons produced in $e^+e^-$ annihilation.

\subsection{Production of back-to-back $\VM$ and $\PS$ mesons}\label{sec:back-to-back VM PS}
It is also interesting to study the qualitative prediction of the string+${}^3P_0$ model for the process $e^+e^-\rightarrow h\, H\, X$, with $h=\VM$ and $H=\PS$ being produced back-to-back in the quark and antiquark jets, respectively. This asymmetry has never been measured. 

The probability $dP(e^+e^-\rightarrow h\,H\,X)$, with $h=\VM$ and $H=\PS$ being produced in the $\q$ and $\qbar$ splittings, can be performed as in Sec. \ref{sec:back-to-back PS PS}. The expression is similar to Eq. (\ref{eq:e+e- -> h H X calculated}), with the second line substituted by the splitting function for $\VM$ emission $F_{\qp,h=\VM,\q}(\mh^2,Z_+,\pt;\kt,\Cqq)\,d\mh^2\,Z_+^{-1}dZ_+\,d^2\pt$. The expression for $F_{\qp,h=\VM,\q}$ is obtained by Eq. (\ref{eq:splitting F_VM explicit}) taking $\Cqq_{x0}=\Cqq_{y0}=0$. Instead, the splitting function for $F_{\qpbar,H,\qbar}(Z_-,\Pt;\ktbar,\C^{\qp\qbar})$ can be obtained from Eq. (\ref{eq:splitting F_PS explicit qbar}). The latter depends on the correlation coefficients $\C^{\qp\qbar}_{0x}$ and $\C^{\qp\qbar}_{0y}$ for $\VM$ emission, which can be calculated using Eqs. (\ref{eq:Cq'qbar})-(\ref{eq:M matrices quark}). Assuming that the decay products of the $\VM$ are not analyzed, we obtain the correlation coefficients
\begin{equation}\label{eq:Cq'qbar calculated VM}
\begin{aligned}
\C^{\qp\qbar}_{0x} &= \C^{\q\qbar}_{xx}\,\frac{M^{\q}_{x0}}{M^{\q}_{00}}= +\frac{\sin^2\theta}{1+\cos^2\theta}\,\fL\,\hat{a}(\kptabs)\,(-\sin\phikptQ), \\
    \C^{\qp\qbar}_{0y} &= \C^{\q\qbar}_{yy}\,\frac{M^{\q}_{y0}}{M^{\q}_{00}}= +\frac{\sin^2\theta}{1+\cos^2\theta}\,\fL\,\hat{a}(\kptabs)\,\cos\phikptQ.
\end{aligned}
\end{equation}
This leads to the probability distribution for $e^+e^-\rightarrow (h=\VM)\, (H=\PS)\, X$ (the analogue of Eq. (\ref{eq:P(e+e- -> h H X) final}))
\begin{align}\label{eq:P(e+e- -> h H X) final VM}
\nonumber        &\frac{dP(e^+e^-\rightarrow h\,H\,X)}{d\cos\theta\,d\mh^2\,dZ_+\,d^2\pt\,dZ_-\,d^2\Pt} \propto \frac{3}{8}\,(1+\cos^2\theta) \\
        &\times \left(1-\frac{\sin^2\theta}{1+\cos^2\theta}\,\fL\,\hat{a}(p_{\rm T})\,\hat{a}(P_{\rm T})\,\cos(\phi^q_h+\phi^q_H)\right).
\end{align}

Comparing with Eq. (\ref{eq:P(e+e- -> h H X) final}), one can see that the Collins asymmetry for back-to-back $\VM$ and $\PS$ mesons has the opposite sign with respect to the asymmetry for back-to-back $\PS$ mesons. Also, it is scaled by the factor $\fL$ and it is thus sensitive to the fraction of longitudinally polarized VMs produced in hadronization. This is a genuine prediction of the string+${}^3P_0$ model that could be tested experimentally, and it is similar to the prediction for the Collins asymmetries for VM production in SIDIS \cite{Kerbizi:2021M20}, and to the prediction for the single spin asymmetries for VM production in $pp$ scattering \cite{Czyzewski:1996ih}.

The high precision BELLE \cite{Belle:2019nve} and BABAR \cite{BaBar:2013jdt} data could provide valuable information on the Collins asymmetry for, e.g., $\rho^0$ and $\pi^{\pm}$ mesons produced back-to-back in $e^+e^-$ annihilation. The measurement of such asymmetry would be useful to retrieve information on the free parameter $\fL$ of the string+${}^3P_0$ model for $\VM$ production. A negative asymmetry would be a confirmation of the prediction of the string+${}^3P_0$ mechanism of hadronization.

\section{Conclusions}\label{sec:conclusions}
We presented an extension of the string+${}^3P_0$ model to the fragmentation of a string stretched between a quark $\q$ and an antiquark $\qbar$ with entangled spin states. The spin correlations of the quarks are described by their joint spin density matrix $\rho(\q,\qbar)$. The latter is calculated assuming the quark pair to be produced in the annihilation of an electron and a positron via the exchange of a virtual photon and neglecting gluon radiation. The model is formulated as a recursive recipe that applies the rules of the string+${}^3P_0$ model for the emission of hadrons from quark splittings as well as the Collins-Knowles recipe to take into account the spin correlations in the fragmentation chain. The recipe is general and it can be applied to other processes as well, regardless of the production mechanism of the $\q\qbar$ pair.

To show that the proposed recipe reproduces the already predicted angular distribution of the final state hadrons, we carried a proof-of-concept calculation for the reaction $e^+e^-\rightarrow \,h\,H\,X$ where one of the hadrons is produced in the quark jet and the other in the antiquark jet. We obtained qualitatively the angular modulation in the distribution of the sum of the azimuthal angles of the two hadrons as expected by the product of two Collins effects. It agrees with the azimuthal correlation observed by the BELLE and BABAR experiments, and predicts a reversal of the sign of the Collins asymmetry for back-to-back pseudoscalar and vector mesons.

For a deeper investigation of the model predictions a Monte Carlo implementation is required. The straightforward choice is the implementation in the PYTHIA generator by extending the StringSpinner package, which will be addressed in a separate work. Other improvements of the model are possible, such as the inclusion of gluon radiation. This development would be important to shed more light on the evolution of quark spin effects with the c.m.s energy of the $e^+e^-$ event.

\begin{acknowledgments}
The authors are grateful to John Collins, Leif Lönnblad and Anna Martin for the many useful discussions. The work of AK is done in the context of the POLFRAG project, CUP n. J97G22000510001, funded by the Italian Ministry of University and Research (MUR).
\end{acknowledgments}

\appendix

\section{Helicity amplitudes}\label{sec:helicity amplitudes}
\begin{figure}[tbh]
\centering
\begin{minipage}[b]{0.48\textwidth}
\includegraphics[width=1.0\textwidth]{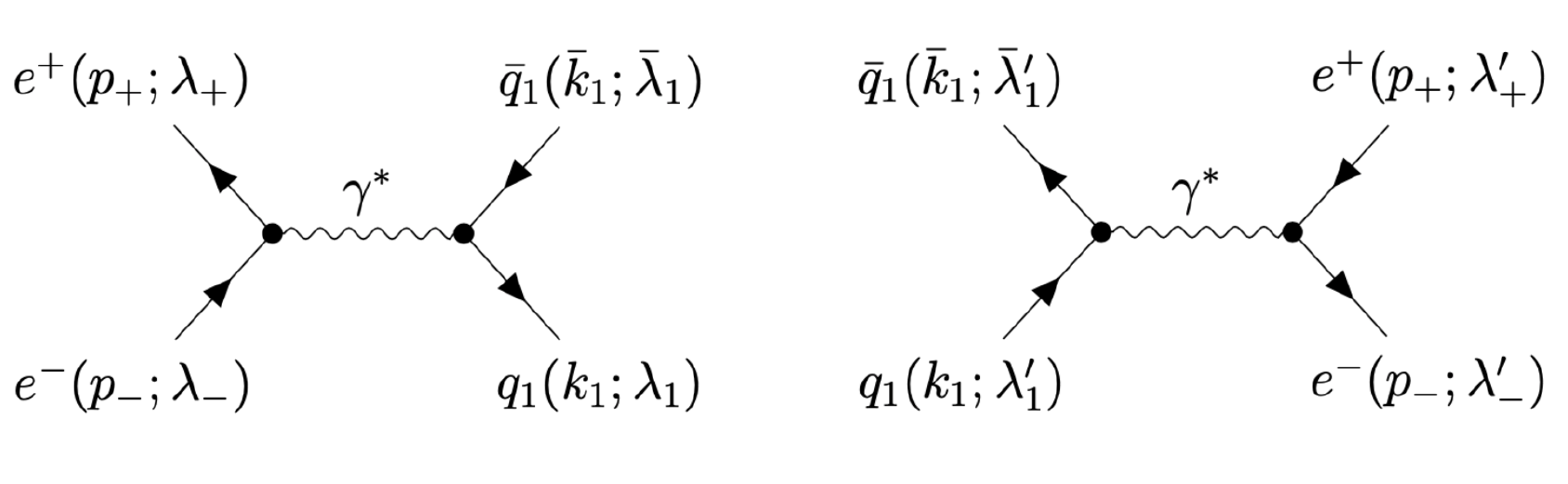}
\end{minipage}
\caption{Leading order diagram for the process $e^+e^-\rightarrow \q\qbar$ (left) and its complex conjugate (right). For each particle the momentum and helicity variables are shown in parenthesis.}
\label{fig:diagram e+e- -> qqbar}
\end{figure}

The Feynman diagram associated to the process $e^+e^-\rightarrow \q\qbar$, considering the exchange of one virtual photon, and its complex conjugate are shown in Fig. \ref{fig:diagram e+e- -> qqbar}. We have indicated by $\helElec$, $\helPos$, $\helQuark$, $\helAntiQuark$ the helicities of the $e^-$, $e^+$, $\q$ and $\qbar$, respectively. The corresponding helicities in the complex conjugated diagram are $\helElec'$, $\helPos'$, $\helQuark'$ and $\helAntiQuark'$. The helicity amplitude associated to the process $e^+e^-\rightarrow \q\qbar$ is given by
\begin{align}\label{eq:M helicity}
\nonumber i\M_{\helElec,\helPos;\helQuark,\helAntiQuark} =i\frac{4\pi\alpha}{s}\,&\left[\bar{v}(e^+,\helPos)\gamma^{\mu}u(e^-,\helElec)\right]\\
&\times\left[\bar{u}(\q,\helQuark)\gamma_{\mu}v(\qbar,\helAntiQuark)\right],
\end{align}
where $u$ and $v$ indicate the Dirac spinors in the helicity basis for a fermion and an anti-fermion, respectively.

The explicit expressions of $\M_{\helElec,\helPos;\helQuark,\helAntiQuark}$ have been calculated keeping the quark mass $m_q$, and the results are shown in Tab. \ref{tab:helicity amplitudes} for the different combinations of the helicity pairs $(\helElec,\helPos)$ and $(\helQuark,\helAntiQuark)$. If the quark masses are neglected, there are only two pairs of non-vanishing matrix elements, i.e. $\M_{+-;+-}=\M_{-+;-+}$ and $\M_{+-;-+}=\M_{-+;+-}$ (the equalities hold because of parity conservation). 

\begin{table}[t]
\begin{tabular}{ |p{1.55cm}|p{1.55cm}|p{1.70cm}|p{1.70cm}|p{1.4cm}|  }
 \hline
 \multicolumn{5}{|c|}{Results for $(4\pi\alpha)^{-1}\,\M_{\helElec,\helPos;\helQuark,\helAntiQuark}$} \\
 \hline
 \hline
 $\,\,\,\,\,\,\,\,\,(\helQuark,\helAntiQuark)$ $\,\,\,\,\,\,\,\,(\helElec,\helPos)$  & $++$ & $+-$ & $-+$ & $--$ \\
 \hline
$+-$ & $\sin\theta/\gamma_q$ & $-(1+\cos\theta)$  & $-(1-\cos\theta)$  & $\sin\theta/\gamma_q$\\
$-+$ & $-\sin\theta/\gamma_q$ & $-(1-\cos\theta)$  & $-(1+\cos\theta)$  & $-\sin\theta/\gamma_q$\\
 \hline
\end{tabular}\caption{The calculated expressions for $(4\pi\alpha/s)^{-1}\,\M_{\helElec,\helPos;\helQuark,\helAntiQuark}$ for the allowed combinations of helicities when taking into account the quark mass and neglecting the electron mass.}\label{tab:helicity amplitudes}
\end{table}

If the quark mass is taken into account, the two other pairs of matrix elements $\M_{+-;++}=\M_{+-;--}\equiv$ and $\M_{-+;++}=\M_{-+;--}$ show up. They involve helicity non-conservation at the quark creation vertex. The explicit expressions are shown in Tab. \ref{tab:helicity amplitudes}. These matrix elements are suppressed by the Lorentz factor $\gamma_q =1/\sqrt{1-\betaq^2}$. For charmed quarks, it is $\gamma_c\simeq 3.57$ in the kinematics of the BELLE and BABAR experiments, and $\gamma_c\simeq 1.20$ and in the kinematics of the BESIII experiment. 

\section{Spin propagation after a hadron emission from the $\qbar$ end}\label{app:VM emission antiquark}
We consider here the case of a hadron $H$ that is emitted in the splitting $\qbar\rightarrow H + \qpbar$ taken from the antiquark end of a string stretched between the quark $\qp$ and the antiquark $\qbar$. The joint spin density matrix of the $\qp\qbar$ system is described by the correlation coefficients $\C^{\qp\qbar}$, calculated using Eq. (\ref{eq:Cq'qbar}).

\paragraph{Density matrix and decay of the VM.} If $H=\VM$, its spin density matrix can be most simply calculated using Eq. (\ref{eq:rho(h) quark}) with the replacement $T^a_{\qp,h=\VM,\q}\otimes\Iqbar \rightarrow \Iq\otimes\,T^{a}_{\qpbar,H=\VM,\qbar}$ and $T^{\dagger\,a'}_{\qp,h=\VM,\q}\otimes\Iqbar \rightarrow \Iq\otimes\,T^{\dagger a'}_{\qpbar,H=\VM,\qbar}$. Using the expression for the splitting amplitude in Eq. (\ref{eq:T antiquark}), the spin density matrix  is
\begin{align}\label{eq:rho(h) antiquark}
 \rho_{aa'}(H)= \frac{\C^{\qp\qbar}_{0\beta}\rm\, \Tr\left[\Delta(\kptbar)\,\Gammaqbar_a(H)\,\sigmaqbar_{\beta}\,\GammaqbarDag_{a'}(H)\,\Delta^{\dagger}(\kptbar)\right]}{\Tr\left[\rm{\dots}\right]}.
\end{align}
The spin density matrix is used to generate the decay of $H$, as described in Sec. \ref{sec:decays}. In this case, however, the polar and azimuthal angles involved in the distribution of the decay hadrons (see Eq. (\ref{eq:decay matrix})) are defined in the AHF. They can be expressed in the QHF by using Eq. (\ref{eq:angles}).

\paragraph{Calculation of the new correlation coefficients.} After the emission of $H$ (and after its decay, if it is a VM), a new string piece stretched between $\qp$ and $\qpbar$ remains. The joint spin density matrix of the $\qp\qpbar$ system can be calculated as in Eq. (\ref{eq:density matrix q'qbar}) using the replacement $T_{\qp,h=\PS,\q}\otimes\Iqbar \rightarrow \Iq\otimes\,T_{\qpbar,H=\PS,\qbar}$ if $H=\PS$, and the replacement $T_{\qp,h=\VM,\q}\otimes\Iqbar \rightarrow \Iq\otimes\,T^{a}_{\qpbar,H=\VM,\qbar}$ if $H=\VM$. The new joint spin desity matrix is thus given by
\begin{equation}\label{eq:density matrix q'qbar'}
\begin{aligned}
    \rho(\qp,\qpbar) &= \begin{cases}
        \T_{\qpbar,H,\qbar} \,\rho(\qp,\qbar)\, \T^{\dagger}_{\qpbar,H,\qbar}/\Tr\left[\rm{\dots}\right] & \rm{H=\PS}\\
        \T^a_{\qpbar,H,\qbar} \,\rho(\qp,\qbar)\, \T^{\dagger\,a'}_{\qpbar,H,\qbar}\,D_{a'a}/\Tr\left[\rm{\dots}\right] & \rm{H=\VM}
    \end{cases}.
\end{aligned}
\end{equation}

The correlation coefficients for the $\qp\qpbar$ system can be calculated using Eq. (\ref{eq:density matrix q'qbar'}) and the explicit expression for the splitting amplitude in Eq. (\ref{eq:T antiquark}). They are 
\begin{align}\label{eq:Cq'qbar'}
   \C^{\qp\qpbar}_{\alpha\beta'} = \Cqq_{\alpha\beta}\,M^{\qbar}_{\beta\beta'}\,/\,\Cqq_{0\beta}\,M^{\qbar}_{\beta0},
\end{align}
where the matrix $M^{\qbar}_{\beta\beta'}$ is given by
\begin{equation}\label{eq:M matrices antiquark}
    \begin{aligned}
        &M^{\qbar}_{\beta'\beta}|_{\PS}  = \frac{1}{2}\, \Tr\left[ \sigmaqbar_{\beta}\,\Delta(\kptbar)\,\Gammaqbar_{H}\,\sigmaqbar_{\beta'}\,\GammaqbarDag_H\,\Delta^{\dagger}(\kptbar)\right]\\
        &M^{\qbar}_{\beta'\beta}|_{\VM} = \frac{1}{2}\,\Tr\left[ \sigmaqbar_{\beta}\,\Delta(\kptbar)\,\Gammaqbar^a(H)\,\sigmaqbar_{\beta'}\,\Gamma^{\dagger\,a'}(H)\,\Delta^{\dagger}(\kptbar)\right]\, D_{a'a}.
    \end{aligned}
\end{equation}
Equations (\ref{eq:Cq'qbar'}) and (\ref{eq:M matrices antiquark}) are, respectively, the analogues of Eqs. (\ref{eq:Cq'qbar}) and (\ref{eq:M matrices quark}) that apply when the meson is emitted from the quark side. The matrix $M^{\qbar}_{\alpha\beta}$ in Eq. (\ref{eq:M matrices antiquark}) can be obtained from Eq. (\ref{eq:M matrices quark}) by the substitutions $\q\rightarrow \qbar$, $h\rightarrow H$ and $\kpt\rightarrow \kptbar$. The matrix $D_{aa'}$ is the decay matrix of $H$, which is calculated as in Sec. \ref{sec:decays}.

\section{Steps for the simulation of the antiquark splitting}\label{sec:steps antiquark splitting}
The steps for the simulation of the antiquark splitting are similar to those for the quark splitting described in the points (2)-(3) in Sec. \ref{sec:recursive recipe}. They are the following:
\begin{itemize}
\item[a.] Select a new quark pair $\qp\qpbar$ with probability $\hat{u}_{\qbar,H}/\hat{u}_{\qbar}$ using Eq. (\ref{eq:u matrix}). Form the hadron $H=(\qbar \qp)$ and decide whether it is a VM with probability $\fVM$, or a PS meson. If $H$ is a PS meson, generate $\kptbarkptbar$, $\phikptQbar$ and $Z_-$ using the splitting function $F_{\qpbar,H=\PS,\qbar}(Z_-,\Pt;\ktbar,\C^{\qp\qbar})$ in Eq. (\ref{eq:splitting F_PS explicit qbar}). If $H$ is a VM use instead the splitting function $F_{\qpbar,H=\VM,\qbar}(\mh^2,Z_-,\Pt;\ktbar,\C^{\qp\qbar})$ in Eq. (\ref{eq:splitting F_VM explicit antiquark}) to generate first the invariant mass squared $\mh^2$, and then $\kptbarkptbar$, $\phikptQbar$ and $Z_-$. 
Calculate the lightcone momenta $P^{-}=Z_-\kbar^-$ and $P^{+}=\ehbar^2/P^-$, and construct the four-momentum of $H$ using $P=(E_H,\Pt,P_L)$, where $E_H=(P^++P^-)/2$ and $P_L=(P^+-P^-)/2$. Express $P$ in the QHF using Eq. (\ref{eq:angles}).
Before accepting the generated $P$, update the new available light-cone momenta $(\Ptot^-)^{\rm new}=\Ptot^- -P^-$ and $(\Ptot^+)^{\rm new}=\Ptot^+ -\ehbar^2/P^-$, and the transverse momentum $\PtotT^{\rm new}=\PtotT-\Pt$. If $\Ptot^2<M_{\rm min}^2$, go to the step 4 in Sec. \ref{sec:recursive recipe}. Otherwise define $\Ptot=\Ptot^{\rm new}$.

If $H$ is a VM apply the following further steps to decay the meson:
    \begin{itemize}
       \item [a.1] Calculate the spin density matrix of $H$ using Eq. (\ref{eq:rho(h) antiquark}), and generate the momenta of the decay hadrons in the rest frame of $H$ using Eq. (\ref{eq:decay distribution}). Express the momenta in the QHF using Eq. (\ref{eq:angles}). The expressions for the decay amplitude $\mathcal{M}$ can be found in Ref. \cite{Kerbizi:2021M20}.
        \item [a.2] Calculate the decay matrix $D$ using Eq. (\ref{eq:decay matrix}).
        \item [a.3] To come back to the center of mass frame, apply the composition of longitudinal and transverse boosts in Ref. \cite{Kerbizi:2021M20} to the decay hadrons. 
    \end{itemize}
    
\item[b.] Calculate the correlation coefficients $\C^{\qp\qpbar}$ of the new string piece with endpoints $\qp$ and $\qpbar$ using Eqs. (\ref{eq:Cq'qbar'})-(\ref{eq:M matrices antiquark}). Let $\qpbar$, $\kpbar$ and $\C^{\qp\qpbar}$ take the place of $\qbar$, $\kbar$ and $\C^{\q\qbar}$, and then go to the step 1 in Sec. \ref{sec:recursive recipe}.
\end{itemize}

\bibliography{bibliography}

\end{document}